\documentclass[3p,times]{elsarticle}
\usepackage{amssymb}
\usepackage{amsbsy}
\usepackage{bm}
\usepackage{latexsym}
\usepackage[active]{srcltx}
\usepackage{latexsym,amsfonts,amsthm,amsmath,amscd,amssymb,color}
\biboptions{sort&compress}

\newcommand{\sect}[1]{\setcounter{equation}{0}\section{#1}}

\numberwithin{equation}{section}

\def\spazio#1{\vrule height#1em width0em depth#1em}

\def\spazio#1{\vrule height#1em width0em depth#1em}

\def\ta{\Bigl(\Bigr.}
\def\tc{\Bigl.\Bigr)}

\begin{document}


\title{\bf {
		Two body relativistic wave equations}}

\author{{R. Giachetti}}
\address{{Dipartimento di Fisica, Universit\`a di Firenze, Italy\\
Istituto Nazionale di Fisica Nucleare, Sezione di Firenze, Italy}}
\author{{E. Sorace}}
\address{{Istituto Nazionale di Fisica Nucleare, Sezione di Firenze, Italy}}

\begin{abstract}
{{		The relativistic quantum mechanics of two interacting
	particles is considered. We first present a covariant formulation of
	kinematics and of reduced phase space, giving
	a short outline of the classical results. We then quantize the
	systems for the scalar-scalar, fermion-scalar and fermion-fermion cases.
	We study the spectrum and the spherical waves solutions
	of the free case. The interaction with central 
	scalar and vector potentials is introduced and the explicit 
	equations are deduced. The one particle and the non 
	relativistic limits are recovered and the general lines for the solution 
	of the boundary value problems are given. We make a numerical analysis of 
	the first two cases with Coulomb interaction. For the two fermions
	we largely revisit the model we
	had previously derived in order to  uniformize the  description for all the three cases.  In order to give 
	a complete review we report in Appendix some of the most interesting 
	results obtained for  atomic and mesonic systems with Coulomb and 
	Cornell potential interactions respectively.}}

 
\end{abstract}


\maketitle

%
%
%
%

\bigskip

%
%
%
\sect{Introduction} \label{Sec_intro}
%
%
%

Just after the formulation of special relativity,
the motion of a single material point, either free or in an external field, was given  
a variational formulation,  both Lagrangian and Hamiltonian  
\cite{Planck_1906,Minkowski_1908,Sommerfeld_1916}.
The difficulties of describing the dynamics of many bodies in the new theoretical framework
were also very soon realized. Even the kinematics of two mass points posed the question of 
the status of the  independent time coordinates. The problem was then dealt with
by reducing the system to a non-relativistic approximation where the energy, given by the sum of 
two square roots, was expanded in powers of $v/c$ and treated by the usual methods of particle dynamics \cite{Darwin_1920}. The rise of Quantum Mechanics shifted the main 
interest towards the search of a relativistically correct quantization procedure. The Dirac 
equation was an essential achievement in this direction and, in addition to countless results in 
fundamental physics, it stimulated many attempts to find an extension for many fermion systems
and their bound states \cite{CVA_1997}. Darwin gave a quantum version of his previous paper
\cite{Darwin_1927}, but the most interesting results were probably produced by Breit 
\cite{Breit_1929}.
Following suggestions independently formulated by Heisenberg and Pauli, Breit assumed as electron 
velocity the Dirac {$\alpha$}-matrices, substituted  them in the Li\'enard-Wiechert potentials 
and gave a first order correction to the Coulomb potential of the Darwin Lagrangian. In this way
he highlighted the strict relationship between the electron motion, 
the spin and the magnetic moment 
as a relativistic feature. Subsequently he  understood that, while the Coulomb potential 
should be treated exactly, the magnetic term had to be taken only at a first perturbation order 
\cite{Breit_1930}.  
In a paper of the same year \cite{Plesset_1932} Plesset  proved a result
contrary to a naif intuition based on a non-relativistic picture: namely he showed that
asymptotically unbounded positive potentials 
in vector coupling  have no  bound states. More detailed descriptions of 
the continuous spectrum for such and similar situations have been given in \cite{GS_2008,GG_2009,GG_2011}.
In fact the covariance properties were not the first
concern of those papers and the equations assumed explicitly  the reference frame with
vanishing total momentum.
Again in the same year, however, a joint paper of Dirac, Fock and Podolsky \cite{DFP_1932}, 
proposed the introduction of  a separate time for each particle to achieve an ``obvious relativistic invariance''. 

A huge breakthrough into the subject was given by a great paper of Wigner \cite{Wigner_1939},
whose analysis of  the relativistically admissible physical states concerned both Quantum Field 
Theory  and Relativistic Quantum  Mechanics.
Some interesting papers were then devoted to the relativistic definition of the center of mass 
\cite{Pryce_1948} and to the localization of elementary states \cite{NW_1949}.
The latter influential paper was published in the special issue for the Einstein 70th birthday: in 
the same volume Dirac made an analysis of the forms in which the interactions could be 
introduced in the relativistic classical Hamitonian mechanics \cite{Dirac_1949},
while the Dirac constraints theory  was published one year later \cite{Dirac_1950}.
A classic paper in this context was  the one by Bakamijan and Thomas \cite{BT_PR92} on relativistic 
particle dynamics. The authors built a two interacting Dirac particles model, in which the two-body 
theory was for the first time coherently based on the algebra of the Poincar\'e generators and a 
canonical separation between global and relative operators was tried, still assuming for the time an {\it{a priori}} different status.
Interesting results were obtained at the same time by Pirenne \cite{Pir_1947} and Berestetski and Landau 
\cite{BL_1949}, who used the  semiclassical Breit approximation and the Schr\"odinger Coulomb wave 
functions  in order to estimate the hyperfine shifts of the parapositronium and of the 
ground  state of the ortopositronium, including the annihilation term.
Few years later, exploiting the QFT advances \cite{Schw_1951,BS_1951}, these shifts  were 
calculated up to the fifth order in the fine structure constant $\alpha$ 
\cite{KK_1952, FM_1955}. 
However, in those 
calculations and in many of the others that followed, the final results were obtained in the 
semiclassical approximation:  a crucial role was played by the value at the origin of the non 
relativistic Coulomb wave function, raising a long lasting question on the opportunity of adopting
some kind of smearing of the corresponding $\delta$-function \cite{RR,BBD}.

In parallel with the investigations of the relativistic few body systems, the interest increased 
also for the analysis of their non-quantum counterpart, mainly based on the group symmetries and on the constraint theory \cite{HST,GDL,Todorov}: 
the hope was to be able to catch the essence of the relationship 
between canonicity and relativistic covariance, in order to have sound guidelines to the quantization 
procedure. An interesting step forward was the so called ``no interaction theorem'' 
\cite{Currie_1963}. It established the impossibility 
for  the relativistic  dynamics of two or more  massive particles 
of being given a Hamiltonian canonical  formulation in explicit covariant  form, 
unless the particles are free. 
The meaning of such a result was made obvious in a Lagrangian context
\cite{GS_1979,BGJS_1984}, proving the impossibility of arbitrary independent time reparametrizations of the world lines of two interacting relativistic particles.
It was then clarified that the angular momentum was a major obstacle to a canonical and explicitly
covariant formulation of the motion and it seemed that quantum field theories only could overcome
this difficulty. However, obtaining bound states in  QFT was - and, in some measure, 
keeps being - a difficult task, because of the perturbative nature of the theory. 
The great and mainly unsolved problems of the wishfully promising  Bethe-Salpeter approach to 
bound states are an example of these obstacles \cite{Nakanishi}. In any case, the QED corrections of
higher and higher order  made on interacting particle models, even when constructed in a  
semi-relativistic framework, have been able to produce very accurate results for the 
levels and the states of simple atoms \cite{Eides,Karshen}.

The situation is even more difficult with the quarkonium models 
\cite{Martin_quarkonium}, used for a long time 
to investigate the spectrum of meson and baryon masses. The first attempt to get at least
semi-quantitative results  dates back from the seventies of last century. The 
Schr\"odinger equation with a potential was initially used in order to fit the charmonium states.
It was soon realized that relativity could not be neglected and the first pioneering work 
\cite{EGKKLY} was followed by many other papers where relativistic effects were added.
Therefore, although generally starting from a Schr\"odinger equation, the models 
have been made more and more relativistic 
by adding perturbative corrections of higher order.
\cite{LSt,Chi,Gro,GodIsg,Isg,LavMcM,TT,BQRTT,GodNap,RKMP,RKMMP,BGEFG,LuScho}.
The quantum chromodynamics (QCD) was also introduced starting from a non-relativistic 
treatment and expanding the interactions up to the second order in $v/c$ \cite{CasLep}.
In such a way an effective theory, called non relativistic QCD or NRQCD, was built and used for the 
lattice and the  continuum calculations \cite{LepThack,BBL}. 
 From the NRQCD another effective theory was then derived, the potential non 
 relativistic QCD or pNRQCD \cite{BPSV},
 which is among the most popular methods for calculating meson spectra and decays 
 \cite{quarkonium,BEHV}. The lattice program has independently grown very consistently
 and is now able to  deduce good estimates of hyperfine splittings  by taking into account  
 relativistic   and  QCD radiative  effects \cite{LW85,DDHH,DDHHH,LPR}.
 
Papers which have proposed a covariant theory are present in literature.  
Many approaches originated from the 
Bethe-Salpeter equation and are connected with field theory: the spectra of the 
corresponding equations are not so easy to compute.
Fewer models, often obtained 
from constrained dynamics, deal with a consistent relativistic quantum 
mechanical description. Extraneous arguments, however,  are sometimes 
introduced, such as the confinement obtained by a cutoff or by a time-like 
potential, a different treatment of the Breit term depending upon the 
component masses, some \emph{ad hoc} contact interaction 
\cite{CVA_1988,CVA_2004,CYW,CSch,Bra,SeCe}.
Those models generally encounter greater difficulties with the 
spectra of mesons formed by different mass quarks and, specially, when 
calculating the masses of the light mesons.
The relativity is obviously essential for  
light quarks and must therefore be treated with the appropriate accuracy. It was also argued  that  the relativistic influence is stronger for 
systems whose components have different masses \cite{Rich}.

Our initial point of view originated from the massive entrance of  differential geometry  
in classical mechanics. The main ideas came from  group actions on symplectic manifolds, 
geometric quantization, reductions of Hamiltonian systems and more 
generally from the use of the induced representations on homogeneous spaces
\cite{Kost,Kir,S_1980,GS_80,GS_81,GRS_LMN_1980,GRS_LMN_1981,GRS_JMP_1981,G_1981,ST_1982}, also
extended to more general and abstract contexts \cite{BEG,BCGST_1992,BCGST_1996,BCGST_1999}.
The analysis of the coadjoint orbits of the Poincar\'e and Weyl  group \cite{Almond} allowed to 
construct
Hamiltonian systems corresponding to group representations, but the most relevant result was 
that the realization of the  particle dynamics on the mass-shell is obtained by a symplectic
reduction of the original phase space \cite{S_1980,GS_80}.
A similar procedure is made possible for a 
system of two particles since the 
relative time variable becomes cyclic and disappears in the reduced space. 
Therefore one can describe 
a canonical consistent relativistic dynamics with a single time coordinate where the interactions are introduced on the reduced phase space by potentials depending upon the  
modulus of the relative position, which is Lorentz invariant 
in the reduced space. Moreover, the 
no-interaction theorem is respected, because the global position coordinate is a 
Newton-Wigner vector  on which there is a well defined action of the kinematic group.

These nice properties suggested the possibility of adopting the two-body relativistic classical dynamics as a starting point to construct a relativistic wave equation for 
a two particle system. The case we started with concerned
two spin one-half particles \cite{GS1}, since most of the elementary and fundamental 
two body bound systems are formed like that. The quantization was realized  
in analogy to  the Dirac procedure for the single particle.
In such a way we deduced a 16-dimensional vector equation reproducing
at the quantum level the same nice properties of the classical systems. 
The wave equation was written in spherical coordinates for dealing with central interactions, determining the states in terms of energy, global angular momentum
and parity. The rotation invariance reduces to eight for each parity the number of independent
radial functions  and since the latter satisfy four algebraic relations, the ultimate boundary value problem 
to be solved has order four. While the two free fermions
can be easily discussed, the addition of central interactions originates a wave equation whose analytical discussion is too much 
demanding and a numerical treatment is almost compulsory.
What can be still analytically proved is the existence and the form of
two different limits: one is the single particle Dirac limit, 
obtained when the mass of one of two
components goes to infinity while the second remains fixed, giving the scale of mass; 
the other one is the Schr\"odinger limit, found by
letting the ratio $v/c$ be vanishing. The former of these limits
constitutes a severe problem for the treatments originating from 
the Bethe-Salpeter equation, which is not able to reproduce it. 
In the  papers \cite{GS2,GS3,BGS_JPB,BGS_PR} we tested the  physical
reliability of our wave equation by numerical calculations. In the area of
atomic physics we determined the hyperfine
shifts of the simple Hydrogen-like atoms with fermionic nucleus. 
Within the high energy framework we calculated 
the masses of the different families of mesons in the  quarkonium model
using the Cornell potential \cite{EGKKLY_1978} and the Breit term as interactions.
Electromagnetic transition rates were also calculated. The results  are generally in excellent agreement with the experimental measures, proving that the correct
inclusion of relativity is able to provide very accurate answers in
contexts of different nature and for energies varying by several orders of magnitude.

In this paper we present a general theory for obtaining the relativistic
wave equations of two particles with spin 0 and 1/2. Besides the two fermion case,
therefore, we determine the equations for two scalars and for a scalar and a  fermion. We  use the geometrical and kinematical properties so far
developed in order to couple the  single particle relativistic wave equation corresponding to each component. The particular features are
 different, but the general method is the same for all the possible cases.
We give for the first time the treatment of the two scalars and scalar-fermion systems, adding a fresh and largely revisited exposition of the two interacting Dirac fermions, uniformized to the general scheme we are developing. As we stated above, at a fundamental level the most diffuse two body systems are formed by two fermions.  Those composed of two scalar
particles are rare  and the corresponding measurements are generally lacking. 
In the realm of the recently developing physics of pionic atoms one
could think of the ion formed by an $\alpha$-particle and a $\pi^-$ meson. Slightly more frequently elementary scalar-fermion systems are met:
for instance the ${}^4{\mathrm{He}}^+$ ion or the (proton-$\pi^-$)
hydrogenic atom, which has recently raised a considerable interest \cite{EXA}.
One could mention also cases, like the deuterium atom, where one of the partners is a vector particle for which the Proca equation should be used.
This last argument and the phenomenological analysis of the two former ones are excluded from this paper and deferred to future research. However, for the sake of completeness and for showing the effectiveness of the approach we have presented, we report in Appendix a choice of the results obtained in \cite{GS3,BGS_JPB,BGS_PR}, concerning the
hyperfine structure of the Hydrogen like atoms and the mass spectrum of mesons in
the quarkonium model. We only give some bare-bone indications of the phenomenological meaning of the data, referring for details to our previous papers. Some final
observations conclude the paper.

\bigskip

%
%
%
\sect{Kinematics and classical systems} \label{Sec_kinematics}
%
%
%

We introduce the basic kinematic definitions that allow to develop a covariant description of two 
relativistic  particles interacting through potentials uniquely dependent upon the difference of 
their coordinates. This means that no external field is present and that the total
momentum is conserved. We call $m_1$ and $m_2$ the particle masses and we assume, without loss in
generality, that $m_1\geq m_2$. The corresponding coordinates and the momenta  will be
$x_{(i)}^\mu,\,p_{(i)}^\mu $, with $i=1,2$, $\mu=0,1,2,3$. The metric tensor $\eta_{\mu\nu}$ has 
the 
usual signature $(+,-,-,-)$, we denote in boldface the 3-vectors and we use units with
$c=1$. We next introduce the conjugate pairs of the so called `global' and `relative' coordinates
\begin{eqnarray}
\Bigl[\,\,
{P^\mu}=p^\mu_{(1)}+p^\mu_{(2)}\,,\quad
{X^\mu}=(1/2)\,\bigl(x^\mu_{(1)}+x^\mu_{(2)}\bigr)\,\,\Bigr]\,,
\qquad
\Bigl[\,\,
{\tilde{q}^{\,\mu}}=(1/2)\,\bigl(p^\mu_{(1)}-p^\mu_{(2)}\bigr)\,,\quad
{\tilde{r}^{\,\mu}}\,=x^\mu_{(1)}-x^\mu_{(2)}\,\,\Bigr]\,.
\label{Coordinate_Globali_Relative}
\end{eqnarray}

As we stated in the introduction, the
relative time and the corresponding relative energy pose a problem of physical 
interpretation, connected with the existence of the `no interaction theorem' 
\cite{Currie_1963,GS_1979}.
The Lagrangian formulation \cite{GS_1979} relates
the absence of interaction to the existence of independent time coordinates of
the particles: it is natural -- and it has been proved -- that interaction determines correlations
among the world lines, incompatible with arbitrary choices of the 
parametrizations in time. 

For the Hamiltonian reduction of the phase space, needed in the present framework, is not
necessary such a complex formalism as in the general case of a Lie group action on
the phase space \cite{G_1981}. Indeed, since the main problem is given by the presence of
relative time and relative energy, it is sufficient to give the explicit canonical 
transformation (whichever the way it has been obtained) turning the relative time into a cyclic variable and then apply the elementary theory of the canonical reduction. 
The relative time  has thus no 
consequences on the development of the dynamics and plays the role of a gauge function to be 
chosen \emph{a posteriori} in order to recover a complete Minkowki description of the world 
lines of the two particles. 
Starting from the two
mass shell conditions written in terms of the coordinates
(\ref{Coordinate_Globali_Relative}) and taking their sum and difference, we have
\begin{eqnarray}
\bigl(P\cdot\tilde{q}\bigr) =(1/2)\,\,(m_1^2-m_2^2),\qquad (1/2)\,\,P^2+2\tilde{q}^2 =m_1^2 + m_2^2.
\label{Ppiumenoq}
\end{eqnarray}
We find it natural to choose as canonical coordinate the variable 
$(P^2)^{-1/2}(P\cdot\tilde{q})$, which is just the relative energy in the 
$\boldsymbol{P}=0$ frame. To complete the set of the new relative canonical
momenta it is therefore coherent to add the relative 3-momentum  
$\boldsymbol{\tilde{q}}$ boosted to the $\boldsymbol{P}=0$ frame. 
It is obvious that this is not a choice of a particular reference frame but only
a definition of a new set of canonical coordinates to which we add the total
4-momentum $P$ in order to exhaust the momentum variables. We finally complete
the canonical transformation by determining the corresponding position variables
by the standard method of the generating function \cite{LL69}. Geometrically the
matrix of the Lorentz transformation $L^{-1}(\boldsymbol{P})$,  which appears in the canonical variables, plays the role of a set of vierbein  
\begin{eqnarray}
\varepsilon_0^\mu(P)=\,\frac{P^\mu}{\sqrt{P^2}}\,~~~~~~~~~~~
\varepsilon_A^\mu(P)=\eta_A^\mu-\displaystyle{\frac{P_A\,\bigl(\,P^\mu
+\eta_0^\mu\sqrt{P^2}\,\bigr)}{\sqrt{P^2}\,\bigl(\,P_0+\sqrt{P^2}\,\bigr)}}\,~~~(\,A=1,2,3\,)
\label{Vierbein}
\end{eqnarray}
satisfying $~\eta_{\mu\nu}\,\varepsilon_\alpha^\mu(P)\,\varepsilon_\beta^\nu(P)\,=
\,\eta_{\alpha \beta}\,~$ and $~\eta_{\alpha \beta}\,\varepsilon_\alpha^\mu(P)\,
\varepsilon_\beta^\nu(P)\,=
\,\eta^{\mu\nu}.~$
The conjugate pairs of canonical variables are:
\begin{eqnarray}
&{}& \Bigl[\, q_0=\varepsilon^\mu_0(P)\, \tilde{q}_\mu,
\phantom{ {P}\,.}r_0=\varepsilon^\mu_0(P)\, \tilde{r}_\mu\,\Bigr]\,,
\qquad
%
\Bigl[\, {q_A}=\varepsilon_A^\mu(P)\, \tilde{q}_\mu,
\phantom{ {P}\,}
{r_A}=\varepsilon_A^\mu(P)\, \tilde{r}_\mu\,\Bigr]\,,
\spazio{1.4}\cr
&{}&  \Bigl[\, {P^\mu}={P^\mu},   
\phantom{ {P},\,\,.}
{Z^\mu}=X^\mu+\Bigl(\,P^2\,\Bigr)^{-1}\,\Bigl(\,(P\cdot\tilde{q})\,\tilde{r}^\mu -(P\cdot\tilde{r})\,\tilde{q}^\mu \,\Bigr)
+\Bigl(\,P^2\Bigl(\,P^0+(P^2)^{1/2}\,\Bigr)\,\Bigr)^{-1}\,W^{\,0\mu}\,\Bigr]
\label{Variabili_Canoniche}
\end{eqnarray}
where the Pauli-Lubanski tensor is defined as
$~W_{\,\mu\nu} = (P^2)^{-1}\, \epsilon_{\,\mu\nu\rho\sigma}\,P^\rho\,W^\sigma,~$ with 
$~W_\mu= \epsilon_{\,\mu\nu\rho\sigma}\,P^\nu\,\tilde{q}^\rho\,\tilde{r}^\sigma.~$

The action of the Poincar\'e group is properly defined: this is due to the  geometrical nature originating the canonical transformation and it is well established by the general theory \cite{G_1981}.  
Moreover $\varepsilon_A^\mu(P)\,W_\mu=\epsilon_{ABC}\,r_B\,q_C \equiv L_A$
is the orbital angular momentum appearing in our previous papers.
Indeed $\emph{\textbf{Z}}\,$ is a Newton-Wigner position vector for a particle of angular momentum $L_A$ and $Z^0$ has the covariance of the time component
of a Lorentz 4-vector. This is necessary if $Z^0$ has to be assumed as  evolution parameter  of the system. Moreover
$\,\emph{\textbf{q}}$ and 
$\emph{\textbf{r}}\,$ are Wigner vectors of spin one with Lorentz invariant modulus
$~q=\bigl(q_A\,q_A\bigr)^{1/2}\,,~$ $~r=\bigl(r_A\,r_A\bigr)^{1/2}.~$
$\,q_0\,$ and $\,r_0\,$ are 
Lorentz invariant also.
The canonical coordinate $Z^\mu$ in (\ref{Variabili_Canoniche}) differs by a term
$(P^2)^{-1/2}\,P^\mu q_0 r_0$ from the expression written in our previous papers.
While the present choice is canonical and derived by elementary methods,
our former one was motivated by a symplectic formulation based on the coadjoint orbits of the Weyl group and contained the above written term involving arbitrary scalar functions  
\cite{Almond, ST_1982}. As long as there are no external interactions the results on the two particle systems are identical. Concerning the quantum mechanics to be developed later on,  we have
$
\exp\,\bigl[\,i\,\bigl(\,p_1 x_1+p_2 x_2\,\bigr)\,\bigr] =\exp\,\bigl[\,i\,\bigl(\,PZ+p_0q_0-{\boldsymbol{q}}\cdot{\boldsymbol{r}}\,\bigr)\,\bigr]
\,
$
while before the term $i\,p_0q_0$ was absent. However that term can presently be fixed to an arbitrary value and produces a completely irrelevant phase factor.
In terms of the new canonical variables, the relations (\ref{Ppiumenoq}) become
\begin{eqnarray}
{P^{\,2}}/{2\,}+2\,q_0^2=2\,q_A q_A+m_{1}^2+m_{2}^2\,,
\qquad\qquad q_0= ({m_{1}^2-m_{2}^2})\,/\,({2\,\sqrt{P^2}})
\label{Energia_Relativa_Libera} 
\end{eqnarray}

which can be solved in $P^2$ yielding, for the total energy,  the result
\begin{eqnarray}
\lambda\,\equiv \,\sqrt{P^2} =
\bigl(\, q_Aq_A+m_{1}^2\,\bigr) ^{1/2}+\bigl( 
\,q_Aq_A+m_{2}^2\,\bigr) ^{1/2}\,
\label{Energia_Totale_Libera}
\end{eqnarray}
So far we have examined a system of two free particles. 
On the reduced phase space it is straightforward to 
introduce a non trivial interaction built in terms of $r_A$ and possibly $q_A$, which produces
a Lorentz covariant dynamics. A simplest choice is to use a potential function in vector, scalar
or even tensor coupling. In the following sections we shall study the first two  choices, dealing 
with atoms and mesons in a quantum mechanical context. For a potential $V(r)$ in the 
vector coupling, the energy (\ref{Energia_Totale_Libera}) is simply modified by the addition of $V(r)$.
Using the spherical coordinates for the relative part, we easily find the radial momentum  $q_r$,
given by
\begin{eqnarray}
\!\!\!\!\!\!\!\!\!q_r= 
\biggl[\,\biggl(\,\frac{(\lambda+V(r))^2-(m_1^2+m_2^2)}{2\,(\lambda+V(r))}\,\biggr)^2 
-\biggl(\,\frac{m_1^2\,m_2^2}{(\lambda+V(r))^2}+\frac{L^2}{r^2}\,\biggr)^2\,\biggr]^{1/2}
\qquad
L^2 = q_\theta^2 +{q_\phi^2}/{\sin^2\theta} 
\label{Impulso_radiale}
\end{eqnarray}

where $L$ is the absolute value of the conserved angular momentum.

Choosing  the Coulomb potential  $~V(r)=-\alpha/r$ and introducing  
$~u(r)=\lambda+V(r),~$
we obtain the equation for the trajectory
\begin{eqnarray}
{{d\theta}/{du}}= 
\bigl({2L}/{\alpha}\bigr)\,
u\,\Bigl(\, u^4-\bigl({2L}/{\alpha}\bigr)^2\,u^2\bigl(\,\lambda-u\,\bigr)^2\,
-2\bigl(m_1^2+m_2^2\bigr)u^2
+\bigl(m_1^2-m_2^2\bigr)^2\,\Bigr)^{-1/2} 
\label{Equazione_Traiettoria_Coulomb}
\end{eqnarray}

It is
integrated in terms of elliptic functions. For equal masses $m_1=m_2=m$ the solution is expressed by elementary functions and reproduces,  for
 $\,L\gtreqless\alpha/2\,$, 
the
elliptic, parabolic and hyperbolic solutions respectively. 

In the following sections we will determine the wave equations for two interacting
relativistic particles in the three cases of scalar-scalar, scalar-fermion and fermion-fermion systems. A unit system with $\,\hbar=c=1\,$ is used. 
We will denote by $\,\sigma_i\,$, $\,i=x,y,z\,$, and 
$\,\sigma_\pm=\sigma_x\pm i\sigma_y\,$ the Pauli matrices. For later convenience 
we also introduce the $\,n\times n\,$ identity matrix $\,{\boldsymbol{\mathrm I}}_n\,$
and the matrices $\,{\boldsymbol{\mathrm I}}_{n,n} 
= {\mathrm{diag}}({\boldsymbol{\mathrm I}}_n,-{\boldsymbol{\mathrm I}}_n)$.

\bigskip

%
%
%
\sect{The wave equation for two scalars} \label{Sec_free_scalar}

The quantization of Hamiltonians including the sum
of two square roots with the possible addition of a potential has been studied in many papers using different mathematical techniques (see, \textit{e.g.}, \cite{VW}). Here and in the following  we
are concerned with particles having a definite internal tensorial structure,
namely scalars and fermions, and with potentials of vector or scalar nature,
respectively coupled to the energy or to the mass. In this section
we begin by studying the quantization of a system of two free scalar particles with masses $m_1$ and $m_2$, $\,m_1\ge m_2$. A vector interaction -- in particular Coulomb -- 
will then be added.
This case is
the easiest from the point ov view of the wave function structure. As we shall see, however, it contains a peculiar feature due to the presence of the Laplace operator.

It has been known since a long time that the  Hamiltonian form of the  wave equation for scalar particles is better obtained by using  a 2-dim formulation for the Klein-Gordon equation    
\begin{eqnarray}
i\,{\partial\Phi}/{\partial t}-H\,\Phi=0\,,
\quad \Phi = {}^{T}\bigl(\phi_1,\phi_2\bigr)
\label{KG_equation_FV}
\end{eqnarray}
where $\,{}^TA$ denotes transposition of a vector or matrix $A$.
In the Feshbach-Villars representation \cite{FV} the Hamiltonian $H$ 
of (\ref{KG_equation_FV}) for a mass $m$ scalar particle is 
\begin{eqnarray}
 H_{FV} = \tau\,{{\boldsymbol p}^2}/{2m} + \sigma_z\,m\,,\qquad\quad p_k = -i\,{\partial}/{\partial x_k},\quad (k=1,2,3), \qquad\quad \tau=\sigma_z+i\sigma_y\,.
\end{eqnarray}
For future convenience we prefer to use a different basis. The matrix $\tau$  is
similar to the Pauli $\,\sigma_-\,$ by the transformation generated by  $T_{\tau} = \sigma_z+\sigma_+$.
The Hamiltonian becomes then
\begin{eqnarray}
H = -\,\sigma_-\,\,\bigl(\,{{\nabla}^2}/{2m}\,\bigr) +(\sigma_z+2\sigma_+)\,m
=\left(\begin{matrix}
m & 2m \spazio{0.8}\cr
-{{\nabla}^2}/{2m} & -m 
\end{matrix}
\,\right)\,.
\label{HKG}
\end{eqnarray}
Assuming a vector coupling with a central potential $V(r)$, in particular a Coulomb potential $\,-\alpha/r\,$, we  define
\begin{eqnarray}
\lambda(r)= \lambda-V(r)\,, \qquad {\mathrm{in}}~~{\mathrm{particular}} \qquad 
\lambda(r)= \lambda+\alpha/r\,.
\label{lambda_r}
\end{eqnarray}
The actual form of the system (\ref{KG_equation_FV}) reads
\begin{eqnarray}
\Bigl(\,{\mathbf{I}}_{2}\,\lambda(r)-H\,\Bigr)\,\Phi(r) =
\left(\begin{matrix}
\lambda(r)-m & -2m \spazio{0.8}\cr
{{\nabla}^2}/{2m} & \lambda(r)+m  
\end{matrix}
\,\right)\,
\left(\begin{matrix}
\phi_1(r) \spazio{0.8}\cr
\phi_2(r)  
\end{matrix}
\,\right)\,=0
\label{SysKG}
\end{eqnarray}
The Klein-Gordon equation is recovered by solving 
in $\phi_2(r)$ the first equation and substituting in the second one.

\subsection{The states and the wave equation.}

The eigenvalue problem for two scalar particles interacting through a
Coulomb-like potential is written in the form
\begin{eqnarray}
\Bigl(\,{\mathbf{I}}_{4}\,\lambda(r)-\bigl(\,H_1\otimes{\mathbf{I}}_{2}\,
+{\mathbf{I}}_{2}\otimes H_2\,\bigr)
\Bigr)\,\Phi(r) =0
\end{eqnarray}
where $H_i$, $i=1,2$, is the Hamiltonian (\ref{HKG}) for the particle with mass $m_i$
and $~ \Phi=\Phi_1\otimes\Phi_2= {}^{T\,}\bigl(\phi_1,\phi_2,\phi_3,\phi_4\bigr)~$
is the state of the system.

We use the canonical variables (\ref{Variabili_Canoniche}). As we said in Section  
\ref{Sec_kinematics} the action of the Lorentz group on the system is well defined: for simplicity and without loss of generality, we can choose the reference system in which the global spatial momentum is vanishing. 
The eigenvalue system of the two scalar particles specifies then to
\begin{eqnarray}
\left(\begin{matrix}
\lambda(r)-m_1-m_2 & -2m_2  & -2m_1 & 0\spazio{0.6}\cr
-{\boldsymbol{q}^2}/{2m_2} & \lambda(r)-m_1+m_2  & 0  & -2m_1 \spazio{0.6}\cr
-{\boldsymbol{q}^2}/{2m_1}  & 0 & \lambda(r)+m_1-m_2  & -2m_2 \spazio{0.6}\cr
0& -{\boldsymbol{q}^2}/{2m_1}  &-{\boldsymbol{q}^2}/{2m_2} &  \lambda(r)+m_1+m_2
\spazio{0.6}
\end{matrix}
\,\right)\,\,
\left(\begin{matrix}
\phi_1({\boldsymbol{r}})\spazio{0.6}\cr
\phi_2({\boldsymbol{r}})\spazio{0.6}\cr
\phi_3({\boldsymbol{r}})\spazio{0.6}\cr
\phi_4({\boldsymbol{r}})\spazio{0.6}
\end{matrix}
\,\right)\,=\,0\spazio{0.8}
\label{SysKG2}
\end{eqnarray}
$-{\boldsymbol{q}}^2 $ being the Laplacian ${\nabla}^2_{\boldsymbol{r}}$ with
respect to the relative coordinate. 
The matrix operator in (\ref{SysKG2}) is not Hermitian, although all the matrix
elements are such. This is  due to the representation
 (\ref{HKG}) of the Klein-Gordon equation,  sharing the same features.   

Let us  look for a reduction of the system (\ref{SysKG2}).
The first line yields directly an algebraic relation 
\begin{eqnarray}
\Bigl(\lambda(r)-m_1-m_2\Bigr)\,\phi_1({\boldsymbol{r}}) -2m_2,\phi_2({\boldsymbol{r}})  
-2m_1\,\phi_3({\boldsymbol{r}})
\label{algKG_1}
\end{eqnarray}
From the second and third lines we derive another algebraic relation
\begin{eqnarray}
\displaystyle m_2\, \bigl( \lambda(r)-m_1+m_2 \bigr) \,\phi_2({\boldsymbol{r}}) -m_1\,  \bigl( \lambda(r)+m_1-m_2 \bigr)\,\phi_3 ({\boldsymbol{r}})=0
\label{algKG_2}
\end{eqnarray}
By (\ref{algKG_1}) and (\ref{algKG_2}) we express $\phi_2({\boldsymbol{r}})$ and $\phi_3({\boldsymbol{r}})$ in terms of $\phi_1({\boldsymbol{r}})$:
\begin{eqnarray}
\phi_2({\boldsymbol{r}}) = \Bigl(4m_2\lambda(r)\Bigr)^{-1}\, \Bigl(\,\bigl(\lambda(r)-m_2\bigr)^2-m_1^2\,\Bigr) \,\phi_1({\boldsymbol{r}})\,,
\qquad
\phi_3({\boldsymbol{r}}) = \Bigl(4m_1\lambda(r)\Bigr)^{-1}\, \Bigl(\,\bigl(\lambda(r)-m_1\bigr)^2-m_2^2\,\Bigr)\,\phi_1({\boldsymbol{r}})\,.
\label{phi2phi3KG2}
\end{eqnarray}
In order to proceed with the reduction process we need a {\it{prolongation}}
of the system \cite{CRC}: this is due to the presence of the
Laplacian in (\ref{SysKG2}) and it is not needed in the cases of scalar-fermion 
and fermion-fermion systems we will treat later on. The prolongation is
obtained by multiplying the matrix  (\ref{SysKG2}) to the left by the diagonal
operator ${\mathrm {diag}}({\boldsymbol{q}}^2,1,1,1)$. Requiring the maintenance
of the Hermiticity of the matrix elements, in the (1,1) place we  take the product
in the  form  $-{\boldsymbol{q}}\,\bigl(\lambda(r)-m_1-m_2\bigl)\,{\boldsymbol{q}}$.
For a Coulomb potential the same result is obtained by using the symmetrized form.

A simple calculation leads to
\begin{eqnarray}
\Bigl(\lambda(r)-m_1-m_2\Bigr)\,{\nabla}^2\phi_1({\boldsymbol{r}})+
{\nabla}\,\lambda(r)\cdot {\nabla}\,\phi({\boldsymbol{r}})
+4m_1m_2\,\Bigl(\lambda(r)+m_1+m_2\Bigr)\,\phi_4({\boldsymbol{r}})=0
\label{prolungo}
\end{eqnarray}
By isolating $\phi_4(r)$ from the second line of (\ref{SysKG2}) and substituting
$\phi_2(r)$ from (\ref{phi2phi3KG2}) we find the final wave equation for the two 
 scalar particles. 
 %
 %
%
 Letting $\phi_1(r)\equiv\phi(r)$ the interacting wave-equation for the two scalars is:
\begin{eqnarray} 
{\nabla}^2\,\phi({\boldsymbol{r}})+\bigl(\,2\,\lambda(r)\,\bigr)^{-1}\,\,
{\nabla}\,\lambda(r)\cdot {\nabla}\,\phi({\boldsymbol{r}})
+\eta^2(r)\,\,\phi({\boldsymbol{r}})=0
\label{KG2V}
\end{eqnarray}
 By specifying (\ref{KG2V}) to the Coulomb potential, so that
 $\,\lambda(r)= \lambda+{\alpha}/{r}\,$, writing the Laplace operator in
spherical coordinates and letting  $\phi({\boldsymbol{r}})=u(r)\,Y_{\ell,m}(\theta,\phi)$, we find the radial equation
\begin{eqnarray}
\biggl(\,\,\frac{d^2}{dr^2}+\Bigl(\,\frac 2r-\displaystyle\frac{\alpha}{2r^2\,\bigl(\,\lambda+{\alpha}/{r}\,\bigr)}\,\Bigr)\,\frac{d}{dr} +\frac 14\, \Bigl(\,\lambda+\displaystyle\frac{\alpha}{r}\,\Bigr)^2+\frac{\bigl(m_1^2-m_2^2\bigr)^2}
{4\,\bigl(\,\lambda+{\alpha}/{r}\,\bigr)^2}
-\displaystyle\frac{m_1^2+m_2^2}{2}-\frac{\ell\,(\ell+1)}{r^2}\,\,\biggr)\,u(r)=0
\label{KG2Coulomb}
\end{eqnarray}
The eigenvalues $\lambda$  give the total energies of the two scalars. When
$\alpha=0$ into (\ref{KG2Coulomb}) we have the free equation whose eigenvalues are
\begin{eqnarray}
\phantom{XXXXXXX}\lambda=\,\pm\,\ta{q_A\,q_A+m_1^2}\tc^{1/2}
\pm\,\ta{q_A\,q_A+m_2^2}\tc^{1/2}\qquad 
(\,\mathrm{all}\,\,\mathrm{possible}\,\,\mathrm{sign}\,\,\mathrm{combinations}\,)
\label{Autovalori_Liberi}
\end{eqnarray}
in agreement with (\ref{Energia_Totale_Libera}).

\subsection{Limits}

It is straightforward to take the limit of (\ref{KG2V}) for $m_1\rightarrow\infty$. Indeed,
recalling that 
\begin{eqnarray}
	\lambda=m_1+m_2+E,
	\label{lambdaE} 
\end{eqnarray}	
when $m_1$ becomes infinite, the limiting equation reduces to
\begin{eqnarray}
{\nabla}^2\,\phi({\boldsymbol{r}})+\Bigl(\,\bigl(\,\lambda'+{\alpha}/{r}\,\bigr)^2-m_2^2\,\Bigr)\,\phi({\boldsymbol{r}})
\label{equalimitKG2}
\end{eqnarray}
with $\lambda'=\lambda-m_1=m_2+E$. Equation (\ref{equalimitKG2}) is thus the Klein Gordon equation for a scalar with mass $m_2$.

The non relativistic limit is found by explicitly reintroducing the speed of light $c$ and rescaling the variables as follows:
\begin{equation}
M\rightarrow Mc^2,\qquad \mu\rightarrow \mu c^2,\qquad \alpha\rightarrow \alpha/c, \qquad r\rightarrow r/c
\label{riscalamenti}
\end{equation}
Taking the limit $c\rightarrow\infty$ of (\ref{KG2Coulomb}), we recover the 
radial Schr\"odinger equation with the
reduced mass $m_R=m_1m_2/(m_1+m_2)$
\begin{eqnarray}
\biggl(\,\frac{d^2}{dr^2}+\frac 2r\,\frac{d}{dr}
+2m_R\,\bigl(\,E+\frac{\alpha}{r}\,\bigr) -\frac{\ell\,(\ell+1)}{r^2}\,\biggr)\,u(r)=0
\label{ESchrodinger}
\end{eqnarray}

\subsection{Solutions}

The solutions of the free equation are given in terms of spherical Bessel functions
$j_\nu(z)=\bigl(\,\pi/2z\,\bigr)^{1/2}\,J_{\nu+1/2}(z),\,$ and read
\begin{eqnarray}
u(r)=A\,j_\ell\,\bigl(k r\bigr)\,, \qquad
 k = \bigl(2\lambda\bigr)^{-1}\,\Bigl[\,\Bigl(\,\lambda^2-\bigl(\,m_1+m_2\,\bigr)^2\,\Bigr)\,
 \Bigl(\,\lambda^2-\bigl(\,m_1-m_2\,\bigr)^2\,\Bigr)\,\Bigr]^{1/2}
 \label{solfreekg2}
\end{eqnarray}

There exists an analytical solution for the interacting equation (\ref{KG2Coulomb}) also.
It is expressed as a combination of confluent Heun functions $H_c$
\cite{Ron,Slav}.
The solution is given by a combinations of the functions
\begin{eqnarray}
\phi_\pm(r) = A\,\exp\,\Bigl[\,{-(\eta/ 2)\,\,({\lambda r}/{\alpha})}\,\Bigr]\,\,
r^{\,(\,{2\beta-1}\,)/\,{4}}\,\,\bigl(\,\lambda r+\alpha\,\bigr)^{\,(\,{2\gamma+1}\,)/\,{4}}\,\,
H_c\,\bigl(\,\eta,\pm\,\beta,\gamma,\delta,\zeta,\,-{\lambda r}/{\alpha}\,\bigr)
\label{SolHeunKG2}
\end{eqnarray}
where $A$ is an integration constant
 and the parameters of $H_c$ are defined as
\begin{eqnarray}
&{}&\eta = \bigl(\,\alpha/\lambda^2\,\bigr)\,\,
\Bigl[\,\Bigl(\,\bigl(\,m_1+m_2\,\bigr)^2-\lambda^2\,\Bigr)\,
\Bigl(\,\lambda^2-\bigl(\,m_1-m_2\,\bigr)^2\,\Bigr)\,\Bigr]^{1/2}
\cr
&{}&\beta=\Bigl(\,1/4+4j(j+1)-\alpha^2\,\Bigl)^{1/2} \qquad\qquad~~~~
\gamma=({2\lambda^2})^{-1}\,\Bigl(\,\lambda^4-4\,\alpha^2\bigl(\,m_1^2-m_2^2\,\bigr)^2\,
\Bigr)^{1/2}\spazio{0.5}\cr
&{}&\delta=-({2\lambda^4})^{-1}\,{\alpha^2}\,\Bigl(\lambda^4-
\bigl(\,m_1^2-m_2^2\,\bigr)^2\,\Bigr)
\qquad\quad\!\zeta=1/8 +{\alpha^2}/2
\label{ParamHc}
\end{eqnarray}
It is clear from (\ref{lambdaE})  that for bound states, $E<0$,
the parameter $\eta$ is real, such as all the other ones. 

In Table I here below we present the results for the lowest levels of two scalar systems
with different mass components, comparing the Schr\"odinger, 
the Klein-Gordon and the two-body results.

\begin{center}
	\begin{small}
		\begin{tabular}{ccrrr}
			\hline\hline
			$(\,n,\ell\,)$ 
			&$m_1/m_2$
			&$E_{\mathrm{Schr}}$\phantom{XXX}
			&$E_{\mathrm{KG}}$ \phantom{XXX}
			&$E_{\mathrm{num}}$\phantom{XXX}
			\\ \hline \spazio{0.2}
			{(1,0)}
			&$1$
			&$-1.331283\,10^{-5}$
			&$-1.331372\,10^{-5}$
			&$-1.331323\,10^{-5}$
			\\  \spazio{0.2}
			{}
			&$100$
			&$-2.636205\,10^{-5}$
			&$2.636381\,10^{-5}$
			&$2.636361\,10^{-5}$
			\\ \hline \spazio{0.2}
		    {(2,0)}
			&$1$
			&$-3.328210\,10^{-6}$
			&$-3.328354\,10^{-6}$
			&$-3.328268\,10^{-6}$
			\\  \spazio{0.2}
			{}
			&$100$
			&$-6.590514\,10^{-6}$
			&$-6.590799\,10^{-6}$
			&$-6.590793\,10^{-6}$
            \\  \hline
			{(2,1)}
			&$1$
			&$-3.328210\,10^{-6}$
			&$-3.328235\,10^{-6}$
			&$-3.328216\,10^{-6}$
			\\  \spazio{0.2}
			{}
			&$100$
			&$-6.590514\,10^{-6}$
			&$-6.590565\,10^{-6}$
			&$-6.590563\,10^{-6}$	
		\\	\hline\hline \spazio{0.4}			
		\end{tabular}
	\end{small}\\
	Table I
\end{center}
In the first column of Table I we indicate the state defined by the principal quantum
 number $\,n\,$ and by the total angular momentum $\,\ell\,$. In the second one the ratio
of the heavier mass $m_1$ to the lighter $m_2$. This latter mass gives the scale of the energies of the columns 3-5, expressed in units of $m_2\,c^2$.
In the third column we give  the Schr\"odinger levels $E_n=-m_R\,c^2\,\alpha^2/2n^2$,
$m_R$ being the classical reduced mass. In the fourth exact result of spectrum of the Klein-Gordon equation, again with the reduced mass, according to the well known expression
\begin{eqnarray}
E_{\mathrm{KG}}(n,\ell)=\bigl(m_R\,c^2\,\bigr)\,\biggl[\,-1+\biggl(\,1+\alpha^2\,
\biggl(\,n-\ell-\frac 12+
\Bigl(\,\bigl(\ell+\frac 12\, \bigr)^2-\alpha^2\,\Bigr)^{1/2}\,\biggr)^{-2}\,\,\biggr)^{-1/2}\,\biggr].
\label{EKG}
\end{eqnarray}
In the last column we give the levels of equation (\ref{KG2Coulomb}). 
The results have been calculated numerically. Indeed, the
spectral theory of the Heun's equation and its polynomial solutions imply
the discussion of a three term recursion relations. The truncation of the series
solution requires to satisfy certain conditions which are not met in (\ref{KG2Coulomb}).
Cases in which such a discussion has been made possible by the special values
assumed by the parameters of the equations are found in \cite{Horta}.

\bigskip

%
%
\sect{The wave equation for a scalar and a fermion} \label{Sec_free_scalar_fermion}
%
%
%
We next study a free system formed by a scalar particle and a fermion
with respective masses $m_S$ and $m_F$, 
The internal geometrical structure of the fermion adds some slight complications to
the simpler treatment of Section 3. The first question to unravel 
concerns the form of the states for which angular momentum and parity are diagonal. In fact the task is not so complicated, due to the scalar nature of one of the components.

Recall first the basic method leading  to the solution of the 
Coulomb problem for the Dirac equation \cite{BLP}. Using the standard representation 
$~\psi={}^{T}\bigl(\,\varphi,\chi\,\bigr)$ for the Dirac spinor, the angular
dependence of the components $\varphi$ and $\chi$ are given by the so called `spherical spinors' $\Omega_{j\ell m}(\theta,\phi)$ where $\ell$ can only assume the values $j+1/2$ or $j-1/2$ 
(and hence $j=\ell\pm 1/2$). The spherical spinors are the following \cite{BLP}:
\begin{eqnarray}
&{}&
\!\!\!\!\!\!\!\!\!\!\!\!\!\!\!\!\!\!
\Omega_{\,\ell+\frac12,\,\ell,\,m}(\theta,\phi) ={}^T\Bigl(\,
\bigl({j+m}\bigr)^{1/2}\,\bigl({2j}\bigr)^{-1/2}\,Y_{\ell,\,m-1/2}(\theta,\phi)\,,~~
\bigl({j-m}\bigr)^{1/2}\,\bigl({2j}\bigr)^{-1/2}\,Y_{\ell,\,m+1/2}(\theta,\phi)
\,\Bigr)
\spazio{1.0}\cr
&{}&
\!\!\!\!\!\!\!\!\!\!\!\!\!\!\!\!\!\!
\Omega_{\,\ell-\frac12,\,\ell,\,m}(\theta,\phi) ={}^T\Bigl(\,
-\,\bigl({j-m+1}\bigr)^{1/2}\,\bigl({2j+2}\bigr)^{-1/2}\,Y_{\ell,\,m-1/2}(\theta,\phi)\,,~~
\bigl({j+m+1}\bigr)^{1/2}\,\bigl({2j+2}\bigr)^{-1/2}\,Y_{\ell,\,m+1/2}(\theta,\phi)
\,\Bigr)
\label{Spinori_Sferici}
\end{eqnarray}
By space inversion $\varphi(\boldsymbol{r})\rightarrow i\varphi(-\boldsymbol{r})$ and  
$\chi(\boldsymbol{r})\rightarrow -i\chi(-\boldsymbol{r})$. 
The parity of spherical spinors is opposite when $|\ell-\ell'|=1$.  Therefore, if the spinor $\psi$ 
has definite parity and $\varphi$ depends upon $\Omega_{j\ell m}$ with
$\ell=j\pm 1/2$, then $\chi$ depends necessarily on $\Omega_{j\ell' m}$, with $ \ell'=j\mp 1/2$.
The general form of the Dirac spinor is then as follows:
\begin{eqnarray}
\psi={}^{T}\Bigl(\,\varphi,\,\chi\,\Bigr)
=  {}^{T}\Bigl(\,a(r)\,\,\Omega_{j\ell m}(\theta,\phi),\,\,\,
b(r)\,\,\Omega_{j\ell' m}(\theta,\phi)\,\Bigr)
\label{Spinore_di_Dirac}
\end{eqnarray}
For the free Dirac equation, letting $p=|\boldsymbol{p}|$, the radial functions are 
\begin{eqnarray}
a(r)=A\,R_{p\ell}(r),\qquad\quad b(r)=B\,R_{p\ell'}(r),\qquad\quad
R_{p\ell}(r)=2\,p^{1/2}\,j_{\ell}(pr),
\label{R_Radiale_di_Dirac}
\end{eqnarray}
 where $\,A,\,$ and $\,B\,$ are integration constants.

\subsection{The states and the wave equation}

Let us switch to the scalar-fermion system  describing the scalar, as in  
Section \ref{Sec_free_scalar}, by the two component formalism and the fermion by the Dirac equation. According to the two component representation (\ref{SysKG}) of the Klein-Gordon equation,  the states of definite 
angular momentum and parity will now be the eight-component vectors. 
For fixed $\,j,\ell,m\,$ the states $\Phi_{\mathrm{I}}$ and ${\Phi}_{\mathrm{II}}$ of opposite parity have the form
\begin{eqnarray}
&{}&
\Phi_{\mathrm{I\phantom{I}}}(\boldsymbol{r}) = {}^T\Bigl(\,\,
a_1(r)\,\,\Omega_{j\ell m}(\theta,\phi),\,\,\,\, 
a_2(r)\,\,\Omega_{j\ell' m}(\theta,\phi),\,\,\,\, 
a_3(r)\,\,\Omega_{j\ell m}(\theta,\phi),\,\,\,\,
a_4(r)\,\,\Omega_{j\ell' m}(\theta,\phi)
\,\,\Bigr)
\spazio{1.0}\cr
&{}&
\Phi_{\mathrm{II}}(\boldsymbol{r}) = {}^T\Bigl(\,\,
b_1(r)\,\,\Omega_{j\ell' m}(\theta,\phi),\,\,\,\, 
b_2(r)\,\,\Omega_{j\ell  m}(\theta,\phi),\,\,\,\, 
b_3(r)\,\,\Omega_{j\ell' m}(\theta,\phi),\,\,\,\,
b_4(r)\,\,\Omega_{j\ell  m}(\theta,\phi)
\,\,\Bigr)
\label{BFStates}
\end{eqnarray}
The explicit form of the Klein-Gordon and Dirac Hamiltonians $H_S$ and $H_F$ of the two particles, after the factorization of the global part and with $q_A=-i\,\partial/\partial r_A\,,~ A=x,y,z,\,$ is
\begin{eqnarray}
H_S = \left(\begin{matrix}
m_S & 2m_S\spazio{1.}\\
{\boldsymbol{q}^2}/{2m_S} & -m_S
\end{matrix}\right)\qquad
H_F = \left(\begin{matrix}
m_F & 0 & q_z & q_x-iq_y \spazio{0.5}\\
0 & m_F & q_x+iq_y & q_z \spazio{0.5}\\
q_z & q_x-iq_y &  -m_F & 0 \spazio{0.5}\\
q_x+iq_y & -q_z & 0 & -m_F 
\end{matrix}\right)
\end{eqnarray}
The global Hamiltonian 
$H=H_S\otimes\mathbf{I}_{4}+\mathbf{I}_{2}\otimes H_F$ thus reads
\begin{eqnarray}
H = \left(\begin{matrix}\spazio{0.8}
m_S+m_F& 0& q_0& \sqrt{2}\,q_-& 2m_S& 0& 0& 0 \spazio{0.5}\\ 
0& m_S+m_F& -\sqrt{2}\,q_+& -q_0& 0& 2m_S& 0& 0\spazio{0.5}\\ 
q_0& \sqrt{2}\,q_-& m_S-m_F& 0& 0& 0& 2m_S& 0\spazio{0.5}\\ 
-\sqrt{2}\,q_+& -q_0& 0& m_S-m_F& 0& 0& 0& 2m_S\spazio{0.6}\\ 
{\boldsymbol{q}^2}/(2m_S)& 0& 0& 0& -m_S+m_F& 0& q_0& \sqrt{2}\,q_-\spazio{0.5}\\ 
0& {\boldsymbol{q}^2}/(2m_S)& 0& 0& 0& -m_S+m_F& -\sqrt{2}\,q_+& -q_0\spazio{0.5}\\ 
0& 0& {\boldsymbol{q}^2}/(2m_S)& 0& q_0& \sqrt{2}\,q_-& -m_S-m_F& 0\spazio{0.5}\\ 
0& 0& 0& {\boldsymbol{q}^2}/(2m_S)& -\sqrt{2}\,q_+& -q_0& 0& -m_S-m_F\spazio{0.5}
\end{matrix}\right)\nonumber\\
\label{HBF}
\end{eqnarray}
In (\ref{HBF}) $\,\,	q_\pm,\,q_0$ are the spherical 
differential operators
\begin{eqnarray}
q_\pm=-( \, \pm{q_{x}} + i\,{q
	_{y}}\,)/\sqrt{2}\,,\qquad\qquad q_0={q_{z}}\,,\qquad\qquad \,q_A \rightarrow 
-i\partial/\partial r_A\,
\label{Xpm0}
\end{eqnarray}
In order to determine the radial functions, we work out explicitly 
the eigenvalue problem relative to the Hamiltonian (\ref{HBF})
for the states (\ref{BFStates}). We obtain a system of equations 
by imposing the vanishing of the coefficients of the different 
spherical harmonics for all the components of the resulting vector.
From the relations thus found, only four linearly independent 
equations can be extracted for each parity. We make the change of variables
$\,\,a_2(r)\rightarrow -ia_2(r)\,\,$ and $\,\,a_4(r)\rightarrow -ia_4(r)\,\,$.
	For the first state (\ref{BFStates}),
$\,\Phi_{\mathrm{I}}\boldsymbol{r})\,$, we have
\begin{eqnarray}
&{}&\bigl(\,{d}/{dr}-r^{-1}\,\bigl(\,j-1/2\,\bigr)\,\bigr) \,a_1(r)
+\bigl(\,\lambda(r)-m_S+m_F\,\bigr)\,a_2(r)-2m_S\,a_4(r)=0\spazio{0.6}\cr 
&{}&\bigl(\,{d}/{dr}+r^{-1}\,\bigl(\,j+3/2\,\bigr)\,\bigr) \,a_2(r)
-\bigl(\,\lambda(r)-m_S-m_F\,\bigr)\,a_1(r)+2m_S\,a_3(r)=0\spazio{0.6}\cr
&{}&\bigl(\,{d}/{dr}+r^{-1}\,\bigl(\,j+3/2\,\bigr)\,\bigr) \,a_4(r)
-\bigl(\,\lambda(r)+m_S-m_F\,\bigr)\,a_3(r)-\bigl(2m_S\bigr)^{-1}\,
{\nabla^2}\,a_1(r)=0
\spazio{0.6}\cr
&{}&\bigl(\,{d}/{dr}-r^{-1}\,\bigl(\,j-1/2\,\bigr)\,\bigr) \,a_3(r)
+\bigl(\,\lambda(r)+m_S+m_F\,\bigr)\,a_4(r)+\bigl(2m_S\bigr)^{-1}\,
{\nabla^2}\,a_2(r)=0 
\label{SysBFeven}
\end{eqnarray}
By a direct computation it can be seen that the equations for the second state 
(\ref{BFStates}), $\,\Phi_{\mathrm{II}}(\boldsymbol{r})\,$, are obtained from (\ref{SysBFeven}) by simply changing  $m_F$ into $-m_F$.

As it occurs in the case of two scalars, here also the system can be reduced because
of the presence of some algebraic relations. The procedure is a bit lengthy but 
straightforward. Isolate $a_4(r)$ from the first and  $a_3(r)$ from the second of
equations  (\ref{SysBFeven}) and substitute in the third and fourth.  When the explicit 
expression of the Laplacian operator is taken into account the second derivatives
cancel and the angular parts contribute with terms  $\,J^2/r^2\,$. The value of the constant $J^2$ is different for the two equations. In the third equation, when calculating $\nabla^2a_1(r)$, we have $J^2=(j-1/2)\,(j+1/2)$; in the fourth one, when calculating $\nabla^2a_2(r)$, $J^2=(j+1/2)\,(j+3/2)$. 
Letting 
$a_1(r)=f(r)$ and $a_2(r)=g(r)$, for the state $\,\Phi(\boldsymbol{r})\,$ we get
the reduced system
\begin{eqnarray}
&{}&
\frac{df(r)}{dr}-\biggl(\,\frac{j-1/2}{r}-\frac {d\lambda(r)/dr}{2\lambda(r)}\,
\,\biggr)\,f(r)
-\biggl(\frac{\lambda(r)}{2}+m_F+\frac{m_F^2-m_S^2}{2\lambda(r)}
\biggr)\,g(r)=0
\spazio{1.2}\cr
&{}&
\frac{dg(r)}{dr}+\biggl(\,\frac{j+3/2}{r}+\frac {d\lambda(r)/dr}{2\lambda(r)}\,
\,\biggr)\,g(r)
+\biggl(\frac{\lambda(r)}{2}-m_F+\frac{m_F^2-m_S^2}{2\lambda(r)}
\biggr)\,f(r)=0
\label{RedSysEven}
\end{eqnarray}
The reduced system for $\,\Phi_{\mathrm{II}}(\boldsymbol{r})\,$ is again found from
(\ref{RedSysEven}) by changing $\,m_F\,$ into $\,-m_F\,$. In particular, for a
Coulomb potential, the system (\ref{RedSysEven}) becomes
\begin{eqnarray}
&{}&\frac{d\,f(r)}{dr}-\frac 1r\,\biggl(\,j-\frac 12+\frac{\alpha}{2\,\bigl(\,\lambda r + \alpha\,\bigr)}\,
\biggr)\,f(r)
-\biggl(\,\frac {\lambda}{2}+m_F +\frac{\alpha}{2r}+\frac{m_F^2-m_S^2}{2\,\bigl(\,\lambda r+\alpha\,\bigr)}\,r\,\biggr)\,g(r)=0\spazio{1.0}\cr
&{}&\frac{d\,g(r)}{dr}+\frac 1r\,\biggl(\,j+\frac 32-\frac{\alpha}{2\,\bigl(\,\lambda r + \alpha\,\bigr)}\,
\biggr)\,g(r)
+\biggl(\,\frac {\lambda}{2}-m_F +\frac{\alpha}{2r}+\frac{m_F^2-m_S^2}{2\,\bigl(\,\lambda r+\alpha\,\bigr)}\,r\,\biggr)\,f(r)=0
\label{KGD_Coulomb}
\end{eqnarray}

\subsection{Limits}

Let us  look at the two separate limits of (\ref{SysBFeven}) for the scalar or the fermion mass tending to infinity. In both cases the first step is to substitute
$\,\,\lambda(r)=m_S+m_F+E-V(r)\,\,$ into (\ref{SysBFeven}).

Consider $m_S\rightarrow\infty$. The first two equations give immediately
$\,a_3(r)=a_4(r)=0$ and the last two ones become identities. Defining 
$\lambda_F(r)= m_F+E-V(r)$, we are therefore left with 
\begin{eqnarray}
&{}&\Bigl(\,{d}/{dr}-r^{-1}\,\bigl(\,j-1/2\,\bigr)\,\Bigr) \,a_1(r)
+\Bigl(\,\lambda_F(r)+m_F\,\Bigr)\,a_2(r)=0\spazio{0.6}\cr
&{}&\Bigl(\,{d}/{dr}+r^{-1}\,\bigl(\,j+3/2\,\bigr)\,\Bigr) \,a_2(r)
-\Bigl(\,\lambda_F(r)-m_F\,\Bigr)\,a_1(r)=0
\label{MBDiracLimit}
\end{eqnarray}
which is exactly the Dirac equation for the spinor (\ref{Spinore_di_Dirac}) with $\,j=\ell+1/2$. The same procedure for the state with opposite parity leads to the
Dirac equation with $\,j=\ell-1/2$.

Take now $m_F\rightarrow\infty$  again in (\ref{SysBFeven}). Substitute the first order expansions $a_k(r) = m_F\,a_{k,1}(r)+a_{k,0}(r)$ for the radial functions and consider the different orders in $m_F$.
From the first and the last equation of (\ref{SysBFeven}) we have
$\,a_{2,1}=a_{4,1}=0$. At order one in $m_F$:
\begin{eqnarray}
&{}&\!\!\!\!\!\!\!\!\!\!\!\!\!\!
a_{2,0}(r)=-\bigl(1/2\bigr)\,\Bigl(\,{d}/{dr}-r^{-1}\,\bigl(\,j-1/2\,\bigr)\,\Bigr) \,a_{1,1}(r)
\qquad
a_{4,0}(r)=-(1/2)\,\Bigl(\,{d}/{dr}-r^{-1}\,\bigl(\,j-1/2\,\bigr)\,\Bigr) \,a_{3,1}(r)
\spazio{0.8}\nonumber\\
&{}&\!\!\!\!\!\!\!\!\!\!\!\!\!\!
a_{3,1}(r) = \bigl(2\,m_S\bigr)^{-1}\,\Bigl(\,E-V(r)\,\Bigr)\,a_{1,1}(r)
\qquad\qquad\qquad
\bigl({\nabla}^2/2\,m_S\bigr)\,a_{1,1}(r)+\Bigl(2\,m_S+E-V(r)\Bigr)\,a_{3,1}(r)=0
\end{eqnarray}
We therefore see that $a_{2,0}(r),\,a_{3,1}(r),\,a_{4,0}(r)$ are all determined by
$a_{1,1}(r)$ which satisfies the Klein-Gordon equation
\begin{eqnarray}
{\nabla}^2\,a_{1,1}(r)+\bigl({\lambda_S}(r)^2-m_S^2\bigr)\,a_{1,1}(r)=0
\label{MFKGLimit}
\end{eqnarray}
where ${\lambda_S}(r)=m_S+E-V(r)$.
Analogous results are obtained for the state  $\,\Phi_{\mathrm{II}}(\boldsymbol{r})\,$.

The non relativistic limit is easily seen to reproduce the Schr\"odinger equation
(\ref{ESchrodinger}).

\subsection{Solutions}

The free wave equation, obtained from (\ref{RedSysEven}) with $\lambda(r)=\lambda=constant$, has the solution
\begin{eqnarray}
f(r)=A\,\,j_j(k r)\qquad\qquad 
g(r)=2A\,\,\Bigl(\,\bigl(\lambda+m_F)^2-m_S^2\,\Bigr)^{-1}\,k{\lambda}\,\,j_{j+1}
(k r)
\end{eqnarray}
where $k$ is as in (\ref{solfreekg2}) with $m_1=m_F,\,\,m_2=m_S$.

The interacting problem does not admit solutions in terms of special functions
but for equal masses, in which case the solution is again a combination of Heun's
confluent functions. In Table II here below we give the numerical results for the
spectrum,
calculated by deriving a second order equation from (\ref{KGD_Coulomb}). The actual
values of the levels are obtained multiplying the data by reduced mass
$~m_Fm_S/(m_F+m_S)$.
\begin{center}
	\begin{small}
		{{ \begin{tabular}{c|r|r|r|r|r}
					\hline\hline
					{{State}}
					&{KG}\phantom{XX}
					&${m_S}/{m_F}=0.1$
					&${m_S}/{m_F}=1$
					&${m_S}/{m_F}=10$
					&{D}\phantom{XXX}
					\\
					\hline					
					$1s_{1/2}$ 
					& -266.274498
					& -266.269982
					& -266.257658
					& -266.258384
					& -266.260317
					\\ 
					$2s_{1/2}$
					& -66.567073
					& -66.566335 
					& -66.564470
					& -66.564894
					& -66.565301
					\\ 
					$2p_{1/2}$  
					& -66.564710
					& -66.564586
					& -66.564470
					& -66.565070
					& -66.565301
					\\ 
					$2p_{3/2}$
					& -66.564710
					& -66.564575
					& -66.564248
					& -66.564337
					& -66.564415
					\\ 
					$3s_{1/2}$
					& -29.585005
					& -29.584764 
					& -29.584184
					& -29.584343
					& -29.584480
					\\ 
					$3p_{1/2}$
					& -29.584305
					& -29.584247
					& -29.584184
					& -29.584396
					& -29.584480
					\\
					$3p_{3/2}$
					& -29.584305
					& -29.584247 
					& -29.584118
					& -29.584178
					& -29.584217
					\\ 
					$3d_{3/2}$
					& -29.584165
					& -29.584148 
					& -29.584118
					& -29.584189
					& -29.584217
					\\ 
					$3d_{5/2}$
					& -29.584165
					& -29.584148 
					& -29.584096
					& -29.584116
					& -29.584130
					\\ 
					\hline\hline
				\end{tabular}
		}}
	\end{small}\\\spazio{1.6}
	Table II
\end{center}
Some comments are in order. Concerning the numerical precision, the data have been calculated in such a way that the figures appearing in the table are all meaningful.
The different properties of level structure with varying mass ratio appear clearly from
the data. Indeed, coherently with what we have shown for infinite mass limits, for 
increasing $m_S/m_F$ we pass from a Klein-Gordon (second column) to a Dirac (last
column) behavior, switching from an approximate degeneracy in $\ell$ to an 
approximate degeneracy in $j$ (columns 3-5). When the level energy increases these degeneracies enhance, so that, at the accuracy presented in the table, the levels appear to be coincident. Instead,
for equal fermion and scalar masses, the degeneracy of the levels with opposite parity is exact. This is due to the fact that the parity is changed by letting
$m_F \rightarrow -m_F$ and for $m_F=m_S$ the resulting second order equations are identical. Finally we see that a term crossings are present exactly at  $m_S/m_F=1$.
This occurs also for the two fermion systems \cite{GS1}. In that case, however, the
hyperfine interaction described by the Breit term removes the crossings. In the
present situation the hyperfine interactions are absent and higher order 
corrections should be considered to see whether crossings survive. 

\bigskip

%
%
\sect{The wave equation for two fermions} 
%
%
%

We finally quantize the system of two relativistic fermions of spin 1/2 in terms of
two coupled Dirac equations. As stated in the Introduction, this is 
the only case we had considered \cite{GS1,GS2,GS3,BGS_JPB,BGS_PR} and
we  present it here because of the ample revisitation of its construction, so to give a
 a coherent and uniform treatment of all the simplest two body relativistic wave equations. 
Being a two body problem, the Hilbert space of the states is given by
the tensor product of the spaces of the two Dirac spinors . We consider the two independent Dirac 
operators in the global and relative coordinates (\ref{Coordinate_Globali_Relative}):
\begin{eqnarray}
{D_1}=\Bigl({P_\mu}/2+\tilde{q}_\mu\Bigr)\,
\tilde{\gamma}_{(1)}^\mu-m_1\,,\qquad
{D_2}=\Bigl({P_\mu}/2-\tilde{q}_\mu\Bigr)\,\tilde{\gamma}_{(2)}^\mu-m_2\,\spazio{1.2}
\label{Dirac1_Dirac2}
\end{eqnarray}
The matrices  $\,\tilde{\gamma}_{(1)}^\mu=\gamma^\mu\otimes{\bf I}_4\,$ and $\,\tilde{\gamma}_{(2)}^\mu={\bf I}_4\otimes\gamma^\mu\,$ operate on each fermion space. Using the canonical coordinates (\ref{Variabili_Canoniche}) and coherently defining
$~{{\gamma}_{(i)}}_0(P)=\varepsilon_0^\mu(P)\,\,\tilde{\gamma}_{(i)\mu}\,,~$ 
$~\gamma_{(i)\,A}(P)=\varepsilon_A^\mu(P)\,\,\tilde{\gamma}_{(i)\,\mu}\,,~$
we have the operator relations
\begin{eqnarray}
(1/2)\,\, \lambda\,{{\gamma}_{(1)}}_0+ 
q_0\,{{\gamma}_{(1)}}_0
-q_A {\gamma_{(1)}}_{A} = m_1\,,  \qquad
(1/2)\,\, \lambda\,{{\gamma}_{(2)}}_0+ 
q_0\,{{\gamma}_{(2)}}_0
-q_A {\gamma_{(2)}}_{A} = m_2\qquad \lambda=\sqrt{P^2}
\label{Dirac_Variabili_Canoniche}
\end{eqnarray}
As long as $\,P\,$ is conserved the gamma matrices can be given the
usual representation.
We solve (\ref{Dirac_Variabili_Canoniche}) in $\lambda$ and $q_0$:
\begin{eqnarray}
{}&{}& {\lambda}=\phantom{{\frac {1}{2}}}\,\Bigl( {{\gamma}_{(1)}}_0\,{\gamma_{(1)}}_{A}-
{{\gamma}_{(2)}}_0\,{\gamma_{(2)}}_{A}\Bigr)\,q_A+{{\gamma}_{(1)}}_0\,m_1+
{{\gamma}_{(2)}}_0\,m_2\,, \qquad
q_0
=\bigl(\,2\,\lambda\,\bigr)^{-1}\,\Bigl(\,m_{1}^2-m_{2}^2\,\Bigr)\,.
\label{Dirac_Due_Fermioni_Liberi}
\end{eqnarray}
The spectrum of the free system is obtained from the first of equations (\ref{Dirac_Due_Fermioni_Liberi}). 
The free relative momentum components  $q_A$ being conserved, the spectrum is formed by the four eigenvalues (\ref{Autovalori_Liberi}) each one of them with multiplicity four.
Defining $M=m_1+m_2$ and $\mu=m_1-m_2$, we see that $\pm M$ 
and  $\pm \mu$ are the eigenvalues of the two fermions at relative rest.
The components of the  spinor tensor product will be reordered so that
in the eigenstate of the system at rest the eigenvalues appear in the order $M,\,-M,\,\mu,\,-\mu$. 
The global angular momentum and parity are conserved. A parity transformation is
the composition of a spatial inversion $\,\,\bm{r}\rightarrow -\bm{r}\,\,$  
and an internal parity  transformation  represented by the matrix $\gamma_0\otimes\gamma_0={{\bf I}}_{8,8}$. We denote by
$\mathrm{I}$ or $\mathrm{II}$ the two opposite parities.

\subsection{The states and the wave equation}

In the usual angular momentum  notations, the sixteen component vector 
states $\,\Psi_{(\mathrm{I,\,II})}\,$ of definite energy, angular momentum $(j,m)$ and parity are chosen to be ordered in
four singlet-triplet multiplets and reads
\begin{eqnarray}
\!\!\!
\Psi_{(\mathrm{I,\,II})}={}^{T\,}\Bigl(\,
\Psi_{(\mathrm{I,\,II)}}^{(M)},\,
\Psi_{(\mathrm{I,\,II})}^{(-M)},\,
\Psi_{(\mathrm{I,\,II})}^{(-\mu)},\,
\Psi_{(\mathrm{I,\,II})}^{(\mu)}
\,\Bigr)\,,\qquad
\Psi_{(\mathrm{I,\,II})}^{(\Gamma)}={}^{T\,}\Bigl(\,
\psi_{({\mathrm{I,\,II}}),\,\,0}^{(\Gamma)}\,,\,
\psi_{({\mathrm{I,\,II}}),\,1_+}^{(\Gamma)}\,,\,
\psi_{({\mathrm{I,\,II}}),\,1_0}^{(\Gamma)}\,,\,
\psi_{({\mathrm{I,\,II}}),\,1_-}^{(\Gamma)}
\,\Bigr),~~ \Gamma=\pm M,\,\pm\mu\,.
\label{Psi_Piu_Meno}
\end{eqnarray}
The  eigenstates of angular momentum are obtained by 
introducing the  `spherical singlets and triplets', which play the same role as the spherical spinors (\ref{Spinori_Sferici}), (\ref{Spinore_di_Dirac}).
The angular part of the singlets is either simply given by the spherical harmonic
$Y^j_{m}(\theta,\phi)$ or vanishing. 
In terms of the usual $\,\langle j_1,m_1,j_2,m_2\,|\,J,M\rangle\,$ Clebsch-Gordan coefficients the triplets are
\begin{eqnarray}
  &{}&\!\!\!\!\!\!\!\!\!\!
 {\boldsymbol{\Omega}}^{(b)}_{j,m}(\theta,\phi)= 
 {}^{T\,}
 \left(\, \Bigl\langle
 j, m-1 ,1,
 1 \,\Bigr|\,
 j, m 
 \Bigr\rangle\, Y^{j}_{m-1}(\theta,\,\phi),\,
 \Bigl\langle
 j, m ,
 1, 0 \,\Bigr|\,
 j, m  
 \Bigr\rangle\, Y^{j}_{m}(\theta,\,\phi),
 \Bigl\langle
 j, m+1 ,
 1, -1 \,\Bigr|\,
 j, m  
 \Bigr\rangle\, Y^{j}_{m+1}(\theta,\,\phi) 
 \,\right) \spazio{1.0}\cr
  &{}&\!\!\!\!\!\!\!\!\!\!
 {\boldsymbol{\Omega}}^{(c)}_{j,m}(\theta,\phi)= 
 {}^{T\,}
 \left(\, \Bigl\langle
 j-1, m-1 ,
 1,1 \,\Bigr|\,
 j, m 
 \Bigr\rangle\, Y^{j-1}_{m-1}(\theta,\,\phi),\,
 \Bigl\langle
 j-1, m ,
 1, 0 \,\Bigr|\,
 j, m  
 \Bigr\rangle\, Y^{j-1}_{m}(\theta,\,\phi),
 \Bigl\langle
 j-1, m+1 ,
 1, -1 \,\Bigr|\,
 j, m  
 \Bigr\rangle\, Y^{j-1}_{m+1}(\theta,\,\phi) 
 \,\right) \spazio{1.0}\cr
   &{}&\!\!\!\!\!\!\!\!\!\!
 {\boldsymbol{\Omega}}^{(d)}_{j,m}(\theta,\phi)= 
 {}^{T\,}
 \left(\, \Bigl\langle
  j+1, m-1 ,1,
  1 \,\Bigr|\,
  j, m 
 \Bigr\rangle\, Y^{j+1}_{m-1}(\theta,\,\phi),\,
 \Bigl\langle
 j+1, m ,
  1, 0 \,\Bigr|\,
  j, m  
 \Bigr\rangle\, Y^{j+1}_{m}(\theta,\,\phi),
 \Bigl\langle
  j+1, m+1 ,
  1, -1 \,\Bigr|\,
  j, m  
 \Bigr\rangle\, Y^{j+1}_{m+1}(\theta,\,\phi) 
 \,\right) \spazio{1.0}\cr
 &{}&	
\end{eqnarray}
Their explicit expressions therefore read
\begin{eqnarray}
&{}&\!\!\!\!\!\!\!\!{\boldsymbol{\Omega}}^{(b)}_{j,m}(\theta,\phi)=
          {}^{T\,}\left(\sqrt{\displaystyle -{\frac {{(j + m)\,(j - m + 1)}}{{2j\,(j+1)}}}}\,
            Y^{j}_{m-1}(\theta,\,\phi),~~
          {\displaystyle \frac 
	        {m}{\,\sqrt{j\,(j+1)}}}\,Y^{j}_{m}(\theta,\,\phi),~~
          {\displaystyle \sqrt{\frac {
			\,{(j + m + 1)\,(j - m)}}{{2j\,(j+1)}}}}\,Y^{j}_{m+1}(\theta,\,\phi)\,\right)
\spazio{1.4}\cr
&{}&\!\!\!\!\!\!\!\!{\boldsymbol{\Omega}}^{(c)}_{j,m}(\theta,\phi)=
          {}^{T\,}\left(\,{\displaystyle \sqrt{\frac {\,{(j + m)\,(j + m - 1)}}
          	 {\,{2j\,(2j - 1)}}}}\,Y^{j-1}_{m-1}(\theta,\,\phi),~~
          {\displaystyle \sqrt{\frac {\,{(j + m)\,(j - m)}}{\,{j\,(2j - 1)}}}}\,
            Y^{j-1}_{m}(\theta,\,\phi),~~
          {\displaystyle \sqrt{\frac {\,{(j - m)\,(j - m - 1)}}{\,{2j\,(2j - 1)}}}}\,
            Y^{j-1}_{m+1}(\theta,\,\phi)\,\right) 
\spazio{1.4}\cr
&{}&\!\!\!\!\!\!\!\!{\boldsymbol{\Omega}}^{(d)}_{j,m}(\theta,\phi)= {}^{T\,}\left(\,  
          {\displaystyle \sqrt{\frac {\,{(j - m + 2)\,(j - m + 1)}}{\,{ (2j+3)\,(2j+2)}}}}
            \,Y^{j+1}_{m-1}(\theta,\,\phi),~~
         -{\displaystyle \sqrt{\frac {\,{(j + m + 1)\,(j - m + 1)}}
		    {\,{ (2j+3)\,(j+1)}}}}\,Y^{j+1}_{m}(\theta,\,\phi),\right.
\spazio{1.4}\cr
&{}&
\left.\phantom{XXXXXXXXXXXXXXXXXXXXXXXXXXXXXXXXXXXXXXXX}		    
{\displaystyle \sqrt{\frac {\,
			{(j + m + 2)\,(j + m + 1)}}{\,{ (2j+3)\,(2j+2)}}}}
          \,Y^{j+1}_{m+1}(\theta,\,\phi)\,\right) 
\label{tripletti}
\end{eqnarray}

Using the previous spherical singlets and triplets, the components $\,\Psi_{\mathrm{I}}^{(\Gamma)}\,$ (\ref{Psi_Piu_Meno})
for the state with the first  parity are 
\begin{eqnarray}
&{}& \!\!\!\!\!\!\!\!\!
\Psi_{\mathrm{I}}^{(M)} = {}^T\Bigl(\,
a_0(r)\,Y_{j,m}(\theta,\phi),\,\,\,\,
b_0(r)\,{\boldsymbol{\Omega}}^{(b)}_{j,m}(\theta,\phi)\,\Bigl)\,\qquad\quad~~~\,
\Psi_{\mathrm{I}}^{(-M)} = {}^T\Bigl(\,
a_1(r)\,Y_{j,m}(\theta,\phi),\,\,\,\,
b_1(r)\,{\boldsymbol{\Omega}}^{(b)}_{j,m}(\theta,\phi)\,\Bigl)
\spazio{0.6}\cr
&{}& \!\!\!\!\!\!\!\!\!
\Psi_{\mathrm{I}}^{(-\mu)} =  {}^T\Bigl(\,
0,\,\,c_0(r)\,{\boldsymbol{\Omega}}^{(c)}_{j,m}(\theta,\phi)+d_0(r)\,{\boldsymbol{\Omega}}^{(d)}_{j,m}(\theta,\phi)\,\Bigl)\,\qquad~~
\Psi_{\mathrm{I}}^{(\mu)} =  {}^T\Bigl(\,
0,\,\,c_1(r)\,{\boldsymbol{\Omega}}^{(c)}_{j,m}(\theta,\phi)+d_1(r)\,{\boldsymbol{\Omega}}^{(d)}_{j,m}(\theta,\phi)\,\Bigl)
\label{Psipiumeno}
\end{eqnarray}
The state of opposite parity  is obtained as
$\,\Psi_{\mathrm{II}} = \bigl(\,\sigma_x\otimes{\boldsymbol{\mathrm{I}}}_8\,\bigr)\,
\Psi_{\mathrm{I}}\,$.

In (\ref{Psipiumeno}) the eight unknown radial functions $a_i(r),\,b_i(r),\,c_i(r),\,d_i(r)\,$, $i=1,2$, replace the functions $a(r)$ and $b(r)$ of \ref{Spinore_di_Dirac}).
Introducing the interaction requires a precise knowledge of its tensorial nature.
For Hydrogen-like atoms the interaction is represented by a Coulomb potential in  vector coupling; for mesons  in the Quarkonium model it is 
described by the Cornell potential, which has a Coulomb-like vector component and a scalar confining linear term.
The general two body interacting wave equation we consider is therefore the following:
\begin{eqnarray}
\Bigl[\,\Bigl(\,
{\gamma}^0_{(1)}{\gamma_{(1)}}_{A}-
{\gamma}^0_{(2)}{\gamma_{(2)}}_{A}\,\Bigr)\,q_A+(1/2)\,\Bigl(\,{\gamma}^0_{(1)}\!
+\! {\gamma}^0_{(2)}\,\Bigr)\,\Bigl(M{{+}}{{\sigma}}
{{{{r}}}}\Bigr)+
(1/2)\,\Bigl(\,{\gamma}^0_{(1)}\!
-\! {\gamma}^0_{(2)}\,\Bigr)\,\mu -\Bigl(\lambda+ {\alpha}/r\,\Bigr)+V_B(r)
\,\Bigr]\,\Psi(\boldsymbol{r})=0.
\label{Equazione_Generale}
\end{eqnarray} 
In the case of Hydrogen-like atoms $\,\sigma=0\,$ and $\,\alpha=Z\,\alpha_{\mathrm{em}}\,$, 
$\,Z\,$ being the atomic number of the nucleus and $\,\alpha_{\mathrm{em}}\,$ the fine structure constant. For the Quarkonium model $\,\sigma\,$ is known as the `string tension' and
$\,\alpha=(4/3)\,\alpha_{\mathrm{S}}\,$, the latter being the strong running coupling constant \cite{PDG}. 
In both cases  we can add the `Breit term' $\,V_B(r)\,$, responsible for the spin-spin interaction, which has to be treated at the first perturbation order:
\begin{eqnarray}
V_B(r)=({{g}/{2r})\,\left(\,{\gamma}^0_{(1)}{\gamma_{(1)}}_{A}
	\,{\gamma}^0_{(2)}{\gamma_{(2)}}_{A}
	+{\gamma}^0_{(1)}{\gamma_{(1)}}_{A}\,{\gamma}^0_{(2)}{\gamma_{(2)}}_{B}\,
	({r_Ar_B}/{r^2})\,\right)}
\label{Termine_di_Breit}
\end{eqnarray}
The eigenvalue problem (\ref{Equazione_Generale}) has the usual form
$~{{\Bigl(\,H_\sigma-(\lambda+\alpha/r)\,{\bf I}_8\,\Bigr)\,\Psi=0}}~$
where
\begin{eqnarray}
H_\sigma =\left(\,
\begin{matrix}
\bigl(\,M+\sigma\,r\,\bigr)\,\, {\bf I}_{\,4,4} & {{\cal H}_0} \spazio{0.8}\cr
{{\cal H}_0}  &- \mu \,\,{\bf I}_{\,4,4}\cr
\end{matrix}
\,\right) \,,
\qquad
 {{{\cal H}_0}} =\left(\,
 \begin{matrix}
 \phantom{-}0_{\phantom{-}}&  \phantom{-}q_+&  \phantom{-}q_0&  
 \phantom{-}q_-&  \phantom{-}0_{\phantom{-}}&  \phantom{-}q_+& 
 \phantom{-}q_0&  \phantom{-}q_- 
 \spazio{0.4}\cr
 - q_-&  \phantom{-}q_0&  \phantom{-}q_-&  \phantom{-}0_{\phantom{-}}&  
 - q_-&  
 - q_0&   - q_- &  \phantom{-}0_{\phantom{-}} 
 \spazio{0.4}\cr
 \phantom{-}q_0&   - q_+&  \phantom{-}0_{\phantom{-}}&  \phantom{-}q_-&  
 \phantom{-}q_0&  \phantom{-}{q
 	_{+}}&  \phantom{-}0_{\phantom{-}}&   - q_- 
 \spazio{0.4}\cr
 -q_+&  \phantom{-}0_{\phantom{-}}&   - q_+&   - q_0&   - q_+
 &  \phantom{-}0_{\phantom{-}}&  \phantom{-}q_+&  \phantom{-}q_0 
 \spazio{0.4}\cr
 \phantom{-}0_{\phantom{-}}&  \phantom{-}q_+&  \phantom{-}q_0&  
 \phantom{-}q_-&  \phantom{-}0_{\phantom{-}}&  \phantom{-}q_+& 
 \phantom{-}q_0&  \phantom{-}q_-  
 \spazio{0.4}\cr
 - q_-&   - q_0&   - q_-&  0&   - q_-
 &  \phantom{-}q_0&  \phantom{-}q_- &  \phantom{-}0_{\phantom{-}} 
 \spazio{0.4}\cr
 \phantom{-}q_0&  \phantom{-}q_+&  \phantom{-}0&  
 - q_-&  \phantom{-}q_0&   - 
 q_+&  \phantom{-}0_{\phantom{-}}&  \phantom{-}q_- 
 \spazio{0.4}\cr
 - q_+&  \phantom{-}0_{\phantom{-}}&  \phantom{-}q_+&  \phantom{-}q_0&   
 - q_+&    
 \phantom{-}0_{\phantom{-}}&   - q_+&   - q_0
 \end{matrix}
 \,\right)
\label{Blocchi_Differenziali}
\end{eqnarray}
and  $\,\,q_\pm,\,q_0$ are the spherical differential operators (\ref{Xpm0}).
The radial eigenvalue  problem is obtained as in Section 4 by 
letting $\,H_\sigma-(\lambda+\alpha/r)\,{\bf I}_8\,$ act on $\Psi_{({\mathrm{I,II}})}$
and imposing the vanishing of the coefficients of the different 
spherical harmonics for all the components of the resulting vector.
Eight linearly independent equations can be extracted from all the relations
thus found.	Introducing the sum and difference notation
$\xi_{\pm}(r) = \xi_+(r) \pm \xi_-(r),~~ (\xi=a,\,b,\,c,\,d)\,\,$ and the operators
\begin{eqnarray}
{\mathrm{D}}_{[\,j_1,\,j_2\,]}=(2j+1)^{-1/2}\,\,j_1^{1/2}\,\,\bigl(\,{{d}/{dr}}
+ {{j_2}/{r}}\,\bigr)\,
\end{eqnarray}
the radial boundary eigenvalue problem for $\,\Psi_{\mathrm{I}}\,$ reduces to the system
\begin{eqnarray}
&{}&  {\mathrm{D}}_{[\,j,\,j+1\,]}\,\,\,
	{a_{+}(r)}
- {\mathrm{D}}_{[\,j+1,\,j+1\,]}\,\,\,
	{b_{-}(r)}
+\,{{c_{0}}(r)}\,\bigl(\mu  + \lambda(r)\bigr)=0~~~~~ 
\spazio{0.6}\cr
&{}& {\mathrm{D}}_{[\,j,\,j+1\,]}\,\,\,
	{a_{+}(r)}
+ {\mathrm{D}}_{[\,j+1,\,j+1\,]}\,\,\,
	{b_{-}(r)}
-\,\,{{c_{1}}(r)}\,\bigl(\mu  - \lambda(r)\bigr)=0~~~~~
\spazio{0.6}\cr
&{}& {\mathrm{D}}_{[\,j+1,\,-j\,]}\,\,\,
	{a_{+}(r)} + {\mathrm{D}}_{[\,j,\,-j\,]}\,\,\,
	{b_{-}(r)}
-\,{{d_{0}}(r)}\,\bigl(\mu  + \lambda(r)\bigr)=0~~~~~ 
\spazio{0.6}\cr
&{}& {\mathrm{D}}_{[\,j+1,\,-j\,]}\,\,\,
{a_{+}(r)} - {\mathrm{D}}_{[\,j,\,-j\,]}\,\,\,
{b_{-}(r)}
+\,{{d_{1}}(r)}\,\bigl(\mu  - \lambda(r)\bigr)=0~~~~~
\spazio{0.6}\cr
&{}&  {\mathrm{D}}_{[\,j,\,-j+1\,]}\,\,\,{c_{+}(r)}
- {\mathrm{D}}_{[\,j+1,\,j+2\,]}\,\,\,
	{d_{+}(r)}
+ \,{{a_{0}}(r)}\,\bigl(M+\sigma\,r - \lambda(r)\bigr)=0~~~~
\spazio{0.6}\cr
&{}& {\mathrm{D}}_{[\,j,\,-j+1\,]}\,\,\,{c_{+}(r)}
- {\mathrm{D}}_{[\,j+1,\,j+2\,]}\,\,\,
{d_{+}(r)}
- \,{{a_{1}}(r)}\,\bigl(M+\sigma\,r + \lambda(r)\bigr)=0~~~~~
\spazio{0.6}\cr
&{}& {\mathrm{D}}_{[\,j+1,\,-j+1\,]}\,\,\,{c_{+}(r)}
+ {\mathrm{D}}_{[\,j,\,j+2\,]}\,\,\,
{d_{+}(r)}
- \,\,{{b_{0}}(r)}\,\bigl(M+\sigma\,r - \lambda(r)\bigr) =0 ~~~~~
\spazio{0.6}\cr
&{}& {\mathrm{D}}_{[\,j+1,\,-j+1\,]}\,\,\,{c_{+}(r)}
+ {\mathrm{D}}_{[\,j,\,j+2\,]}\,\,\,
{d_{+}(r)}
- \,\,{{b_{1}}(r)}\,\bigl(M+\sigma\,r + \lambda(r)\bigr) =0 ~~~~~
\label{Le_8_Pari}
\end{eqnarray}
The system for $\Psi_{\mathrm{II}}$ is obtained by means of the parity transformation. It can be seen that this amounts to substituting 
$M+\sigma\,r\rightarrow -\mu$ and $\mu\rightarrow -(M+\sigma\,r)$.
	
The differential order for both parity systems is actually four due 
to four algebraic relations among the unknown functions. 
These are more simply written by introducing the following linear combinations: 
\begin{eqnarray}
&{}&\!\!\!\!\!\!\!\! {u_+}(r)=-\,s_0\,{c_+}(r) + 
s_1\,{d_+}(r)\,,
\qquad {u_-}(r)=-\,s_0\,{c_-}(r) + 
s_1\,{d_-}(r)\,,\spazio{0.6}\cr
&{}&\!\!\!\!\!\!\!\! {v_+}(r)=-\,s_1\,{c_+}(r) - 
s_0\,{d_+}(r)\,,
\qquad {v_-}(r)=-\,s_1\,{c_-}(r) - 
s_0\,{d_-}(r)\,,
\label{Variabili_u_v}
\end{eqnarray}
where we have defined $\,s_0 = j^{1/2}\,(2\,j+1)^{-1/2}\,$ and $\,s_1=(j+1)^{1/2}\,(2\,j+1)^{-1/2}\,$. 
The algebraic relations for the state $\Psi_{\mathrm{I}}$ are
\begin{eqnarray}
&{}& a_-(r) = \lambda(r)^{-1}\,\,\bigl(M+\sigma\,r\,\bigr)\,\, a_+(r)
\qquad\quad u_-(r)=- \lambda(r)^{-1}\,\Bigl(\,
 \bigl(\,j(j+1)\,\bigr)^{1/2}\,(2/r)\, b_-(r)-\mu\,u_+(r)\,\Bigr)
\spazio{0.6}\cr
&{}& b_+(r) = \lambda(r)^{-1}\,\,\bigl(M+\sigma\,r\,\bigr)\,\, b_-(r)
\qquad\quad v_+(r)=\phantom{-}\lambda(r)^{-1}\,\Bigl(\,
\bigl(\,j(j+1)\,\bigr)^{1/2}\,(2/r)\, a_+(r)-\mu\,v_-(r)\,\Bigr)\,.
\label{Relazioni_Algebriche_Pari}
\end{eqnarray}
The analogous relations for $\Psi_{\mathrm{II}}$ are again obtained by substituting 
$M+\sigma\,r\rightarrow -\mu$ and $\mu\rightarrow -(M+\sigma\,r)$ in
(\ref{Relazioni_Algebriche_Pari}).
	
	We choose the independent unknown functions $\,\bigl(\,a_+(r),\,\,b_-(r),\,\,u_+(r),\,\,v_-(r)\,\bigr)\,$ which we 
	respectively arrange  in a vector
	$\,\,Y(r)\equiv{}^{T\,}\!\bigl(\,y_1(r),\,y_2(r),\,y_3(r),\,y_4(r)\,\bigr)\,$.
	When $Y(r)$ is known, the state $\,\Psi_{\mathrm{I}}\,$ is reconstructed by the inverse transformation. 
The analogous procedure applies to the odd parity.
The  eigenfunctions for two fermions are thus obtained by solving 
	\begin{eqnarray}
	\left(
	\begin{matrix}
	y'_1(r)\spazio{0.6}\cr
	y'_2(r)\spazio{0.6}\cr
	y'_3(r)\spazio{0.6}\cr
	y'_4(r)\spazio{0.6}\cr
	\end{matrix}
	\right)
	+
	\left(
	\begin{matrix}
	0 & E(r,\lambda) & -F(r,\lambda) & 0 \spazio{0.6}\cr
	E(r,\lambda) & 1/r & 0 & F(r,\lambda) \spazio{0.6}\cr
	G(r,\lambda) & 0 & 2/r & E(r,\lambda) \spazio{0.6}\cr
	0 & -G(r,\lambda) & E(r,\lambda) & 1/r\spazio{0.6}\cr
	\end{matrix}
	\right)\,\,
	\left(
	\begin{matrix}
	y_1(r)\spazio{0.6}\cr
	y_2(r)\spazio{0.6}\cr
	y_3(r)\spazio{0.6}\cr
	y_3(r)\spazio{0.6}\cr
	\end{matrix}
	\right)
	=0
	\label{Le_4_Int}
	\end{eqnarray}
With the usual notation (\ref{lambda_r}) for the Coulomb potential, the matrix elements for the two parities are
\begin{eqnarray}
&{}&\!\!\!\!\!\!\!\!\!\!\!\!\!\!\!\!\!\!\!\!\!\!\!\!\!\!\!
E_{\mathrm{I}\,}(r,\lambda) = \frac{\bigl(\,j\,(j+1)\,\bigr)^{1/2}\,\mu}{r\,\lambda(r)}\,,\qquad\qquad~~~
F_{\mathrm{I}\,}(r,\lambda) = \frac{\lambda(r)}2-\frac{\mu^2}{2\,\lambda(r)}\,,
\qquad~~
G_{\mathrm{I}\,}(r,\lambda) = \frac{\lambda(r)}2-\frac{2j\,(j+1)}{r^2\lambda(r)}-    
    \frac{(M+\sigma r)^2}{2\,\lambda(r)}\spazio{1.2}\cr
&{}&\!\!\!\!\!\!\!\!\!\!\!\!\!\!\!\!\!\!\!\!\!\!\!\!\!\!\! 
E_{\mathrm{II}\,}(r,\lambda) = -\frac{\bigl(\,j\,(j+1)\,\bigr)^{1/2}\,(M+\sigma\,r)}{r\,\lambda(r)}\,,~~ 
F_{\mathrm{II}\,}(r,\lambda) = \frac{\lambda(r)}2-\frac{(M+\sigma\,r)^2}{2\,\lambda(r)}\,,~~
G_{\mathrm{II}\,}(r,\lambda) = \frac{\lambda(r)}2-\frac{2j\,(j+1)}{r^2\lambda(r)}-    
\frac{\mu^2}{2\,\lambda(r)}
\label{EvenOddCoeff} 	
	\end{eqnarray} 
	
\subsection{Limits}
	
Let us look at two limiting cases of (\ref{Le_4_Int}). We first consider the
limit for $m_1\rightarrow\infty$, or `Dirac limit'. In this case 
and with $\,\lambda_2(r)=m_2+E+\alpha/r,\,$,
the coefficient functions of (\ref{Le_4_Int}) for the first parity become
\begin{eqnarray}
&{}& \!\!\!\!\!\!\!\!\!
E^D_{\mathrm{I}\,}(r,\lambda) = \bigl(\,j\,(j+1)\,\bigr)^{1/2}\,/{r}\,,\qquad 
F^D_{\mathrm{I}\,}(r,\lambda) = \lambda_2(r)+m_2\,,\qquad
G^D_{\mathrm{I}\,}(r,\lambda) = \lambda_2(r)-\bigl(m_2+\sigma\,r\bigr)
\spazio{0.8}\cr
&{}& \!\!\!\!\!\!\!\!\!
E^D_{\mathrm{II}\,}(r,\lambda) =-A^D_{\mathrm{e}}(r,\lambda)\,,\qquad\quad\quad~\,\,
F^D_{\mathrm{II}\,}(r,\lambda) =C^D_{\mathrm{e}}(r,\lambda)\,,\quad\quad\quad G^D_{\mathrm{II}\,}(r,\lambda) =B^D_{\mathrm{e}}(r,\lambda)\
\label{EvenOddCoeffDiracLimit}   
\end{eqnarray}
%
The limiting equations obtained by using (\ref{EvenOddCoeffDiracLimit}) 
in (\ref{Le_4_Int}) are equivalent  to pairs of Dirac equations. 
A mixing is  necessary in order to decouple the fourth order system. 
In the two parity cases,  with $\,s_0,\,s_1\,$ as above,
 the mixing transformations are respectively generated by the $2\times 2$ block matrices
\begin{eqnarray}
T_{\mathrm{I}}=\left( 
\begin{array}{cc}
0  & s_0\sigma_z+s_1\sigma_x\cr
s_1{\mathbf{I}}_2 -is_0\sigma_y & 0
\end {array}
\right)
\qquad
T_{\mathrm{II}}=\left( 
\begin{array}{cc}
0  & s_0{\mathbf{I}}_2 +is_1\sigma_y\cr
s_1\sigma_z+s_0\sigma_x & 0
\end {array}
\right)
\end{eqnarray}
In both cases $({y}_1(r),{y}_4(r))$ decouple from $({y}_2(r),{y}_3(r))$ and give Dirac equations
\begin{eqnarray}
{\displaystyle\frac{d}{dr}}\,
\left({\begin{array}{c} r \,f(r) \spazio{0.4}\\ r \,g(r)
	\end{array}}\right)
+
\left({\begin{array}{cc} {\kappa}/{r} 
	& -	\bigl(\,\lambda_2(r)+m_2 \,\bigr) \spazio{0.4}\\
	\lambda_2(r)-m_2-\sigma\,r & -{\kappa}/{r}
	\end{array}}\right)\,
\left({\begin{array}{c} r \,f(r) \spazio{0.6}\\ r \,g(r)
	\end{array}}\right)
\label{Sistema_di_Dirac}
\end{eqnarray}
For the first parity, taking the unknown functions $\bigl(f(r),g(r)\bigr)$ equal to $\bigl(r\,y_4(r),r\,y_1(r)\bigr)$ or to $\bigl(r\,y_3(r),-r\,y_2(r)\bigr)$ we get
(\ref{Sistema_di_Dirac}) with $\kappa=-(j+1)$ or $\kappa=j$, the two values corresponding to orbital angular momentum $\ell=j$.
In the odd case for $\bigl(f(r),g(r)\bigr)$ equal to $\bigl(r\,y_1(r),-r\,y_4(r)\bigr)$ or $\bigl(r\,y_2(r),r\,y_3(r)\bigr)$, we get $\kappa=-j$ or $\kappa=j+1$, the two values 
corresponding to $\ell=j+1$  and $\ell=j-1$ respectively \cite{BLP}.

In order to recover  the Schr\"odinger equation we reintroduce explicitly the
factor $c$. The limit is
obtained from (\ref{Le_4_Int})-(\ref{EvenOddCoeffDiracLimit}) using the rescaling (\ref{riscalamenti}) together with $\sigma\rightarrow\sigma c$.
Eliminating $y_3(r)$ and $y_4(r)$  we get a system of two second order 
differential equations for $y_1(r)$ and $y_2(r)$. When $c\rightarrow\infty$ 
both $\,y_1(z)\,$ and $\,y_2(z)\,$ satisfy the Schr\"odinger equation
\begin{equation}
\Bigl(\,\frac{d^2}{dr^2}+\frac 2r\,\frac{d}{dr}
 +2m_R\,\bigl(\,E-\sigma\,r+\frac{\alpha}{r}\,\bigr)-\frac {j\,(j+1)}{r^2}\,\,\Bigr)\,u(r)=0
\label{ESchrodinger_sigma}
\end{equation}
For the first parity we have two decoupled equations with $\ell=j$ in both cases.
For the second parity, a transformation generated 
by the matrix $\,T=(j+1)^{1/2}\,{\bold{I}}_2+i\,j^{1/2}\,\sigma_y\,$
is needed and produces two equations  (\ref{ESchrodinger_sigma}) with $\ell=j\pm 1$.

\subsection{Solutions}
\smallskip

Analytical solutions are available for the free system (\ref{Le_4_Int})-(\ref{EvenOddCoeff})
with $\alpha=\sigma=0$. They are expressed in terms of spherical Bessel functions.
With integration constants $A$, $B$ and $k$ as in 
(\ref{solfreekg2}),   for parity I we have:
\begin{eqnarray}
&{}& \!\!\!\!\!\!\!\!\!
y_1(r)= \phantom{-}A\,j_{j}\bigl(kr\bigr)	
\qquad\quad
y_2(r)= \phantom{-}B\,\,j_{j}(kr)
\spazio{0.6}\cr
&{}&\!\!\!\!\!\!\!\!\! 
y_3(r)=\phantom{-}\bigl(\lambda^2-\mu^2\bigr)^{-1}\,r^{-1}\,
\Bigl(\,{2A\,j\lambda+2B\,{j^{1/2}\,(j+1)^{1/2}}\,\,\mu}\,\Bigr)
\,\,j_j\bigl(kr\bigr)
-\bigl(\lambda^2-\mu^2\bigr)^{-1}\,{2\,A\,k\lambda}\,\,\,j_{j+1}(kr)
\spazio{0.6}\cr
&{}&\!\!\!\!\!\!\!\!\! 
y_4(r)=-\bigl(\lambda^2-\mu^2\bigr)^{-1}\,r^{-1}\,
\Bigl(\,{2A\,{j^{1/2}\,(j+1)^{1/2}}\,\,\mu+2B\,(j+1)\lambda}\,\Bigr)
\,\,\,j_j(kr)
+\bigl(\lambda^2-\mu^2\bigr)^{-1}\,
{2\,B\,k\lambda}\,\,\,j_{j+1}(kr)
\label{y_Pari_Liberi}
\end{eqnarray}
The solutions for $\Psi_{\mathrm{II}}$ are obtained by substituting 
$\,M\rightarrow -\mu$ and $\mu\rightarrow -M\,$ in
(\ref{y_Pari_Liberi}).

The interacting models  have been discussed numerically. 
The analysis is conducted within the spectral theory in Hilbert spaces by the 
`double shooting method'
with a very high accuracy obtained by  Pad\'e approximants. The matching of the
four components of the system at a crossing point 
$\,0<r_c<\infty\,$ determines a linear system of four 
algebraic equations whose determinant constitutes the spectral condition.. 

For the sake of completeness we reorganize and report in Appendix some results obtained in (\cite{GS3,BGS_JPB,BGS_PR}) for atom and meson systems.


\bigskip

%
%
%
\sect{Conclusions} \label{Sec_em}
%
%
%

Relativistic quantum mechanics has generally been considered a preliminary step, naturally and
unavoidably leading to the quantization of fields. The history of physics demonstrates 
that this idea has produced the fantastic achievements obtained during
the whole last century  and, undoubtedly, keeps being very fruitful. There are however
some subjects that are better dealt with directly in a quantum mechanical framework
rather than in a field theoretical one. This happens in the study 
of the bound states of composite relativistic systems  
when a high degree of accuracy is required. It is not only the case of high 
energy objects, as the hadrons, but also that of atoms, due to the higher and higher
precision recently reached by the measurements. The development of relativistic
quantum mechanics for an arbitrary number of interacting bodies, however, 
has met some problems for which a complete and convincing solution,
provided it  exists, has not yet been given. In particular, difficulties have always been
found in reconciling the covariance with the evolution in time if an interaction
is present. This is well explained by Wigner (1969), when he says: 
`{\emph{It appears reasonable to state that the problem of the motion of several interacting relativistic particles in classical and quantum 
mechanics has not been solved. In constructing such a theory one encounters relative time
coordinates whose meaning is often obscure}}' (quoted by Cook \cite{Cook}). 
We have given our point of view on the restricted case of systems composed of two
particles only, by presenting a procedure that allows to
reduce the phase space  so to make the relative time a 
cyclic variable. We must admit that for two bodies the reduction procedure works fine, but unfortunately
 we do not have the general solution when the number of the particles is larger.
In the restricted case  we have given a covariant treatment of the
problem both at a classical and at a quantum level and we have tested it,
with positive confirmation,  against the
experimental data taken on physical objects of different nature and ranging on 
different scales of energy. The quantum system of two interacting particles
is defined by the total angular momentum, the parity and the invariant mass, which
is the eigenvalue of the Hamiltonian, while the masses of the free fermions enter as
independent parameters. In this way we have built a composite object such that each of its
eigenstates corresponds to a representation of the Poincar\'e group.
The comparison  with the experimental data has implied the recourse to
a numerical analysis of our model. We believe that a more general and deeper mathematical investigation,
in analogy to what has been done for the Dirac equation \cite{Thaller} should be
worthwhile. The main interest could come from the fact that we are studying a 
fourth order problem: this, indeed, presents aspects that are not always
intuitive, due to the commonly rooted habit of reasoning in terms of second order 
-- and mainly elliptic --
equations. From a more physical point of view, we believe that the method  could be generalized to decays \cite{BGS_JPB,BGS_PR} and to include the higher order corrections
of the fields. The two body equation would then provide a better starting point
for obtaining a higher accuracy in the phenomenological analysis of physical systems and
in the comparison with experimental data.

\bigskip

%
%
%
\section{Appendix} \label{App}
%
%

\subsection{Atoms}
\smallskip

The current theoretical calculations found in literature  reproduce rather well 
the measured quantities. The agreement with data improves by  adding effects of different nature by perturbation expansions, generally starting from a non-relativistic description of the physical system. Here we report our data  exclusively
obtained  from equation (\ref{Le_4_Int}) and the Breit term
(\ref{Termine_di_Breit}) treated at the first perturbation order.
The constants are fixed at their measured physical values:
$m_1$ and $m_2$ the masses of the two particles, $\,b=\alpha = Ze^2/\hbar c$,
$e$ being the electron charge, $g=\kappa_p\kappa_e\alpha$ where
$\kappa_1$ and $\kappa_2$ are the factors accounting for the anomalous magnetic moments of the two fermions. The latter are assumed to be point-like, neglecting finite radius corrections even for massive nuclei as ${}^{3}{\mathrm{He}}^+$.
 The model is therefore very sharp, without any free parameter and with a 
completely clear physical content. The results show that the complete inclusion of covariance properties and spin-spin interactions, even without further QED corrections, produce by themselves very high accuracy. The Lamb shift can be introduced in an effective way (we did it in \cite{BGS_JPB}), but it is not considered here, since it gives very very small corrections to the de data we present in Table III, concerning 
the hyperfine splittings (HFS) of the levels. 
\smallskip

\begin{center}
	\begin{small}
	\begin{tabular}{lrr|rr}
		\hline\hline
		{{Atom}} 
		&\phantom{}$\Delta({1s})_{\mathrm{Th}}$\phantom{iii}
		&\phantom{}$\Delta({1s})_{\mathrm{Exp}}$\phantom{iii}
		&\phantom{}$\Delta({2s})_{\mathrm{Th}}$\phantom{iii} 
		&\phantom{}$\Delta({2s})_{\mathrm{Exp}}$\phantom{iii}
		\\
		\hline\spazio{0.2}
		({{p}},$\,${{e}})
		&\phantom{i}{1420.595}\phantom{i}\phantom{$\frac{i}{i}$}
		&\phantom{i}{1420.405}\phantom{i}\phantom{$\frac{i}{i}$}
		&\phantom{i}{177.580}\phantom{i}\phantom{$\frac{i}{i}$}
		&\phantom{i}{177.557}\phantom{i}\phantom{$\frac{i}{i}$}
		\\ \spazio{0.2}
		({$\mu^+,\,\mathrm{e}$})
		&\phantom{i}{4464.481}\phantom{i}\phantom{$\frac{i}{i}$}
		&\phantom{i}{4463.302}\phantom{i}\phantom{$\frac{i}{i}$}
		&\phantom{i}{558.078}\phantom{i}\phantom{$\frac{i}{i}$}
		&\phantom{i}{558.\phantom{000}}\phantom{i}\phantom{$\frac{i}{i}$}
		\\\spazio{0.2} 
		({${}^{3}{\mathrm{He}}^+,$}$\,${{e}})
		&\phantom{i}{-8665.637}\phantom{i}\phantom{$\frac{i}{i}$}  
		&\phantom{i}{-8665.650}\phantom{i}\phantom{$\frac{i}{i}$}
		&\phantom{i}{-1083.347}\phantom{i}\phantom{$\frac{i}{i}$}  
		&\phantom{i}{-1083.355}\phantom{i}\phantom{$\frac{i}{i}$} 
		\\ 
		\hline\spazio{0.2}
		({{p}},$\,${{$\mu$}})
		&\phantom{i}{182.621}\phantom{i}\phantom{$\frac{i}{i}$}
		&\phantom{i}{182.638}\phantom{i}\phantom{$\frac{i}{i}$} 
		&\phantom{i}{22.828}\phantom{i}\phantom{$\frac{i}{i}$}
		&\phantom{i}{22.815}\phantom{i}\phantom{$\frac{i}{i}$} 
		%
		\\\spazio{0.2} 
		(${}^{3}{{\mathrm He}}^+,$ $\,${{$\mu$}})
		&\phantom{i}{-1372.194}\phantom{i}\phantom{$\frac{i}{i}$}
		&\phantom{i}{-1334.730}\phantom{i}\phantom{$\frac{i}{i}$}
		&\phantom{i}{-171.544}\phantom{i}\phantom{$\frac{i}{i}$}
		&\phantom{i}{-166.645}\phantom{i}\phantom{$\frac{i}{i}$} 
		\spazio{0.5}\\
		\hline\spazio{0.4}
%
			{{Atom}} 
			&\phantom{ii}$\Delta({2p}^{1/2})_{\mathrm{Th}}\,$ 
			&\phantom{ii}$\Delta({2p}^{1/2})_{\mathrm{Exp}}\,$  
			&\phantom{ii}$\Delta({2p}^{3/2})_{\mathrm{Th}}\,$
			&\phantom{ii}$\Delta({2p}^{3/2})_{\mathrm{Exp}}\,$  
			\\
			\hline\spazio{0.2}
			({{p}},$\,${{e}})
			&\phantom{i}{59.196}\phantom{i}\phantom{$\frac{i}{i}$}
			&\phantom{i}{59.221}\phantom{i}\phantom{$\frac{i}{i}$}
			&\phantom{i}{23.678}\phantom{i}\phantom{$\frac{i}{i}$} 
			&\phantom{i}{24.\phantom{000}}\phantom{i}\phantom{$\frac{i}{i}$}
			\\\spazio{0.2} 
			({$\mu^+,$}$\,${{e}})
			&\phantom{i}{186.252}\phantom{i}\phantom{$\frac{i}{i}$}
			&\phantom{i}{187.\phantom{000}}\phantom{i}\phantom{$\frac{i}{i}$}
			&\phantom{i}{74.629}\phantom{i}\phantom{$\frac{i}{i}$} 
			&\phantom{i}{74.\phantom{000}}\phantom{i}\phantom{$\frac{i}{i}$}
			\\\spazio{0.2} 
			({${}^{3}{He}^+,$}$\,${{e}})
			&\phantom{i}{-361.100}\phantom{i}\phantom{$\frac{i}{i}$}
			&\phantom{i}{-\phantom{000}}\phantom{i}\phantom{$\frac{i}{i}$}
			&\phantom{i}{-144.385}\phantom{i}\phantom{$\frac{i}{i}$}
			&\phantom{i}{-\phantom{000}}\phantom{i}\phantom{$\frac{i}{i}$}
			\\ 
			\hline\spazio{0.2}
			({{p}},$\,${{$\mu$}})
			&\phantom{i}{7.682}\phantom{i}\phantom{$\frac{i}{i}$}
			&\phantom{i}{7.820}\phantom{i}\phantom{$\frac{i}{i}$}
			&\phantom{i}{3.115}\phantom{i}\phantom{$\frac{i}{i}$}
			&\phantom{i}{3.248}\phantom{i}\phantom{$\frac{i}{i}$}
			\\ \spazio{0.2}
			({${}^{3}{\mathrm{He}}^+,$}$\,${{$\mu$}})
			&\phantom{i}{-57.028}\phantom{i}\phantom{$\frac{i}{i}$}
			&\phantom{i}{-58.713}\phantom{i}\phantom{$\frac{i}{i}$}
			&\phantom{i}{-22.700}\phantom{i}\phantom{$\frac{i}{i}$}
			&\phantom{i}{-24.291}\phantom{i}\phantom{$\frac{i}{i}$}
			\\ 
			\hline\hline
		\end{tabular}
	\end{small}\\ \spazio{1.6}
Table III
	\end{center}
        The first column of Table III indicates the simple atom which we refer to.
		The following columns respectively give our  theoretical and 
		experimental results for the HFS of levels indicated on top.  
		According to the common use,
		when the electron is present the results are given in MHz.
		When the muon is the lightest fermion, the results are in meV. The values of the 
		proton, electron and muon mass,
		$\,m_e=5.485799091\cdot10^{-4}\,u\,$, $\,m_p=1.007276466879\,u\,$, 
		$\,m_{{}^{3}{\mathrm{He}}^+}=3.0160293\,u\,$ (in unified atomic mass unit),
		 have been taken from CODATA 2010. Moreover $\kappa_p=g_p/2=2.7928473565$, $\kappa_e=g_e/2=1.0011596522$, 
		$\kappa_\mu=g_\mu/2= 1.0011659207$.
		For $\,{}^{3}{\mathrm{He}}^+\,$ we have assumed  
		$\kappa=-3.1839627379413$,  obtained taking into account the Helion 
		atomic number, from $\kappa_p$, the ratio  $2.9931526707$ of the Helion to the proton mass and the ratio $-0.761766558$ of the shielded Helion to
		proton magnetic moment. The value of the fine structure constant, is
		$\alpha=0.0072973525698$. Theoretical results obtained by different methods for the HFS of the $s$ and $p$ 
		levels of simple Hydrogen-like atoms are found in the papers 
		\cite{Jungmann,Karshenboim,Haensch_2005,Haensch_2006,Martynenko,Rafel}.
		More details on our results are found in \cite{BGS_JPB}.

\subsection{Mesons}
\smallskip

We have considered the meson mass spectrum in the Quarkonium model using the Cornell 
potential, formed by a Coulomb-like term $\alpha/r$ and a confining scalar term
$\sigma r$. $\sigma$ is called the `string tension', while $\alpha=(4/3)\,\alpha_S\,$ where $\alpha_S$ is the strong running coupling constant. Both these parameters are fixed at a constant value for each meson family. The spin dependent interactions are modeled  by a Breit perturbation term.
\smallskip
\begin{center}
\begin{small}
	{{ \begin{tabular}{c|lrr|lrr|lrr}
				\hline\hline
				{{State}}
				&$b\bar{b}$ 
				&${\mathrm{Exp}}\phantom{XXX}$ 
				&${\mathrm{Num}}$
				&$c\bar{c}$
				&${\mathrm{Exp}}\phantom{XXX}$ 
				&${\mathrm{Num}}$
				&$s\bar{s}$
				&${\mathrm{Exp}}\phantom{XXX}$
				&${\mathrm{Num}}$
				\\
				\hline
				$({{1}}^{{1}}{{s}}_{{0}})~0^+(0^{-+})~
				\phantom{{}^{{}^{\displaystyle{i}}}}\!\!$ 
				&$\eta_b$
				&{9390.90$\pm$2.8}  
				&{9390.39}
				&$\eta_c$
				&{2978.40$\pm$1.2}
				&{2978.26}
				&
				&$\cdot\cdot\cdot\cdot\cdot$\phantom{X}\phantom{iii}
				&{818.12}
				\\ 
				$({{1}}^{{3}}{{s}}_{{1}})~0^-(1^{--})~
				\phantom{{}^{{}^{\displaystyle{i}}}}\!\! $
				&$\varUpsilon$
				&{9460.30$\pm$.25}  
				&{9466.10}
				&$J/\psi$
				&{3096.92$\pm$0.1}
				&{3097.91}
				&$\phi$
				&{1019.46$\pm$.02}
				&{1019.44}
				\\ 
				$({{1}}^{{3}}{{p}}_{{0}})~0^+(0^{++})~
				\phantom{{}^{{}^{\displaystyle{i}}}}\!\!$  
				&$\chi_{b0}$
				&{9859.44$\pm$.73}  
				&{9857.41}
				&$\chi_{c0}$
				&{3414.75$\pm$.31}
				&{3423.88}
				&
				&$\cdot\cdot\cdot\cdot\cdot$\phantom{X}\phantom{iii}
				&{1206.44}
				\\ 
				$({{1}}^{{3}}{{p}}_{{1}})~0^+(1^{++})~
				\phantom{{}^{{}^{\displaystyle{i}}}}\!\! $
				&$\chi_{b1}$
				&{9892.78$\pm$.57}  
				&{9886.70}
				&$\chi_{c1}$
				&{3510.66$\pm$.07}
				&{3502.83}
				&$f_{1,(1420)}$
				&{1426.40$\pm$.90}
				&{1412.84}
				\\ 
				$({{1}}^{{1}}{{p}}_{{1}})~0^-(1^{+-})~
				\phantom{{}^{{}^{\displaystyle{i}}}}\!\! $
				&$h_b$
				&{9898.60$\pm$1.4}  
				&{9895.35}
				&$h_c$
				&{3525.41$\pm$.16}
				&{3523.67}
				&
				&$\cdot\cdot\cdot\cdot\cdot$\phantom{X}\phantom{iii}
				&{1458.59}
				\\ 
				$({{1}}^{{3}}{{p}}_{{2}})~0^+(2^{++})~
				\phantom{{}^{{}^{\displaystyle{i}}}}\!\! $
				&$\chi_{b2}$
				&{9912.21$\pm$.57}  
				&{9908.14}
				&$\chi_{c2}$
				&{3556.20$\pm$.09}
				&{3555.84}
				&$f'_{1,(1525)}$
				&{1525$\pm$\phantom{.}5\phantom{0}}
				&{1525.60}
				\\
				$({{2}}^{{3}}{{s}}_{{1}})~0^-(1^{--})~
				\phantom{{}^{{}^{\displaystyle{i}}}} \!\!$
				&$\varUpsilon$
				&{10023.26$\pm$.0003}  
				&{10009.04}
				&$\psi$
				&{3686.09$\pm$.04}
				&{3692.91}
				&$\phi$
				&{1680$\pm$\phantom{.}20}
				&{1698.41}
				\\ 
				$({{1}}^{{3}}{{d}}_{{2}})~0^-(2^{--})~
				\phantom{{}^{{}^{\displaystyle{i}}}}\!\! $
				&$\varUpsilon_2$
				&{10163.70$\pm$1.400}  
				&{10152.69}
				&$\psi_2$
				&$\cdot\cdot\cdot\cdot\cdot$\phantom{X}\phantom{iii}
				&{3833.62}
				&
				&$\cdot\cdot\cdot\cdot\cdot$\phantom{X}\phantom{iii}
				&{1838.72}
				\\ 
				$({{2}}^{{3}}{{p}}_{{0}})~0^+(0^{++})~
				\phantom{{}^{{}^{\displaystyle{i}}}}\!\! $
				&$\chi_{b0}$
				&{10232.50$\pm$.0009}  
				&{10232.36}
				&$\chi_{c0}$
				&$\cdot\cdot\cdot\cdot\cdot$\phantom{X}\phantom{iii}
				&{3898.00}
				&
				&$\cdot\cdot\cdot\cdot\cdot$\phantom{X}\phantom{iii}
				&{1841.12}
				\\ 
				$({{2}}^{{3}}{{p}}_{{1}})~0^+(1^{++})~
				\phantom{{}^{{}^{\displaystyle{i}}}}\!\! $
				&$\chi_{b1}$
				&{10255.46$\pm$.0005}  
				&{10256.58}
				&$\chi_{c1}$
				&$\cdot\cdot\cdot\cdot\cdot$\phantom{X}\phantom{iii}
				&{3961.21}
				&
				&$\cdot\cdot\cdot\cdot\cdot$\phantom{X}\phantom{iii}
				&{1988.38}
				\\
				$({{2}}^{{3}}{{p}}_{{2}})~0^+(2^{++})~
				\phantom{{}^{{}^{\displaystyle{i}}}}\!\! $
				&$\chi_{b2}$
				&{10268.65$\pm$.0007}  
				&{10274.26}
				&$\chi_{c2}$
				&{3927.00$\pm$2.5}
				&{4003.93}
				&$f_{2,(2010)}$
				&{2011$\pm$\phantom{.}70}
				&{2073.15}
				\\
				\hline\hline
			\end{tabular}
	}}
\end{small}\\\spazio{1.6}
Table IV
\end{center}
In Table IV we report in MeV the masses of the heavy mesons formed by Bottom, Charm and Strange quarks, 
        ${b}{\bar{b}}$,  ${c}{\bar{c}}$ and ${s}{\bar{s}}$.
		The first column contains term symbol and $I^G(JPC)$ numbers. In the last three columns we give the particle 
		name, the experimental data and our results in MeV for  ${b}{\bar{b}}$,  
		${c}{\bar{c}}$ and ${s}{\bar{s}}$ respectively. The masses of the bottom,
		charm and strange quarks, in MeV, are: $m_b=4725.5,\,m_c=1394.5.5,\,m_s=134.27$.
		For ${b}{\bar{b}}$ and ${c}{\bar{c}}$ the string tension is 
		$\sigma$=${1.111}\,$GeV/fm; for ${s}{\bar{s}}$, $\sigma$=${1.34}\,$GeV/fm. 
		Moreover $\alpha_{\mathrm{S}}$=0.3272, 0.435, 0.6075 respectively. 
		\cite{PDG}.

\begin{center}\begin{small}
{{ $~~~$
		\begin{tabular}{lrr|lrr}
\hline\hline
$~~~~~~$State$\phantom{{}^{{}^{\displaystyle{i}}}}$ & Exp\phantom{XX} 
& Num\phantom{Xi}	 &	$~~~~~~$State$\phantom{{}^{{}^{\displaystyle{i}}}}$ & 
Exp\phantom{XX}	 & 
Num\phantom{Xi}		
\\
\hline	
$({{1}}^{{1}}{{s}}_{{0}})~0^+(0^{-+})~
\phantom{{}^{{}^{\displaystyle{i}}}}\!\! ${{{$\pi^\pm$}}}
&\phantom{i}{139.57$\pm$.00035}\phantom{i}  
&\phantom{i}{616.45}\phantom{i}

&	$({{1}}^{{1}}{{s}}_{{0}})~0(0^{-})~
\phantom{{}^{{}^{\displaystyle{i}}}}\!\! ${{{B${}_{{s}}^{{0}}$}}}   
& \phantom{i}{5366.77$\pm$.24}\phantom{i} 
&\phantom{i}{5387.41}\phantom{i}

\\
$({{1}}^{{3}}{{s}}_{{1}})~1^+(1^{--})~
\phantom{{}^{{}^{\displaystyle{i}}}}\!\! ${{{$\rho$(770)}}}
&\phantom{i}{775.49$\pm$.39}\phantom{i}  
&\phantom{i}{\phantom{1}826.14}\phantom{i}

& $({{1}}^{{3}}{{s}}_{{1}})~0(1^{-})~
\phantom{{}^{{}^{\displaystyle{i}}}}\!\! ${{{B${}_{{s}}^{{*}}$}}} 
& \phantom{i}{5415.40$\pm$2.1}\phantom{i} 
&\phantom{i}{5434.34}\phantom{i}

\\
$({{1}}^{{3}}{{p}}_{{0}})~1^-(0^{++})~
\phantom{{}^{{}^{\displaystyle{i}}}}\!\!${{{a${}_{{0}}$(980)}}}
&\phantom{i}{980.$\pm$20}\phantom{i}  
&\phantom{i}{970.34}\phantom{i}

& $({{1}}^{{3}}{{p}}_{{1}})~0(1^{+})~
\phantom{{}^{{}^{\displaystyle{i}}}}\!\!${{{B${}_{{s1}}$(5830)${}^{{0}}$}}} 
& \phantom{i}{ 5829.40$\pm$.70}\phantom{i} 
&\phantom{i}{5817.80}\phantom{i}

\\
$({{1}}^{{3}}{{p}}_{{1}})~1^-(1^{++})~
\phantom{{}^{{}^{\displaystyle{i}}}}\!\!  ${{{a${}_{{1}}$(1260)}}}
&\phantom{i}{1230.$\pm$.40}\phantom{i}  
&\phantom{i}{1204.66}\phantom{i}

& $({{1}}^{{3}}{{p}}_{{2}})~0(2^{+})~
\phantom{{}^{{}^{\displaystyle{i}}}}\!\!${{{B${}_{{s2}}$(5840)${}^{{0}}$}}} 
& \phantom{i}{5839.70$\pm$.60}\phantom{i}  
&\phantom{i}{5829.33}\phantom{i}

\\
\cline{4-6}
$({{1}}^{{1}}{{p}}_{{1}})~1^+(1^{+-})~
\phantom{{}^{{}^{\displaystyle{i}}}}\!\!  ${{{b${}_{{1}}$(1235)}}}
&\phantom{i}{1229.5$\pm$3.2}\phantom{i}  
&\phantom{i}{1274.76}\phantom{i}

& \phantom{i}$({{1}}^{{1}}{{s}}_{{0}})~0(0^{-})~
\phantom{{}^{{}^{\displaystyle{i}}}}\!\! ${{{D${}_{{s}}^\pm$}}} 
& \phantom{i}{1968.49$\pm$.32}\phantom{i} 
&\phantom{i}{1961.24}\phantom{i}

\\
$({{1}}^{{3}}{{p}}_{{2}})~1^-(2^{++})~
\phantom{{}^{{}^{\displaystyle{i}}}}\!\!  ${{{a${}_{{2}}$(1320)}}}
&\phantom{i}{1318.3$\pm$.60}\phantom{i}  
&\phantom{i}{1325.40}\phantom{i}

&  \phantom{i}$({{1}}^{{3}}{{s}}_{{1}})~0(1^{-})~
\phantom{{}^{{}^{\displaystyle{i}}}}\!\! ${{{D${}_{{s}}^{*\pm}$}}}
& \phantom{i}{2112.30$\pm$.50}\phantom{i}  
&\phantom{i}{2101.78}\phantom{i}

\\
$({{2}}^{{1}}{{s}}_{{0}})~1^-(0^{-+})~
\phantom{{}^{{}^{\displaystyle{i}}}}\!\!  ${{{$\pi$(1300)}}}
&\phantom{i}{1300$\pm$100}\phantom{i}  
&\phantom{i}{1337.36}\phantom{i}

&\phantom{i}$({{1}}^{{3}}{{p}}_{{0}})~0(0^{+})~
\phantom{{}^{{}^{\displaystyle{i}}}}\!\!${{{D${}_{{s0}}$(2317)${}^\pm$}}}
&\phantom{i}{2317.80$\pm$.6\phantom{0}}\phantom{i}   
&\phantom{i}{2339.94}\phantom{i}

\\
$({{2}}^{{3}}{{s}}_{{1}})~1^+(1^{--})~
\phantom{{}^{{}^{\displaystyle{i}}}} \!\! ${{{$\rho$(1450)}}}
&\phantom{i}{1465$\pm$\phantom{.}25}\phantom{i}  
&\phantom{i}{1497.63}\phantom{i}

& \phantom{i}$({{1}}^{{3}}{{p}}_{{1}})~0(1^{+})~
\phantom{{}^{{}^{\displaystyle{i}}}}\!\!${{{D${}_{{s1}}$(2460)${}^\pm$}}} 
&\phantom{i}{2459.60$\pm$.6\phantom{0}}\phantom{i}  
&\phantom{i}{2466.15}\phantom{i}

\\
$({{1}}^{{3}}{{d}}_{{1}})~1^+(1^{--})~
\phantom{{}^{{}^{\displaystyle{i}}}}\!\!  ${{{$\rho$(1570)}}}
&\phantom{i}{1570${}^{(*)}$}\phantom{i}  
&\phantom{i}{1565.42}\phantom{i}

& \phantom{i}$({{1}}^{{1}}{{p}}_{{1}})~0(1^{+})~
\phantom{{}^{{}^{\displaystyle{i}}}}\!\!${{{D${}_{{s1}}$(2536)${}^\pm$}}}
&\phantom{i}{2535.12$\pm$.13}\phantom{i}   
&\phantom{i}{2535.82}\phantom{i}

\\
$({{3}}^{{1}}{{s}}_{{0}})~1^-(0^{-+})~
\phantom{{}^{{}^{\displaystyle{i}}}} \!\! ${{{$\pi$(1800)}}}
&\phantom{i}{1812$\pm$\phantom{.}12}\phantom{i}  
&\phantom{i}{1882.30}\phantom{i}

&\phantom{i}$({{1}}^{{3}}{{p}}_{{2}})~0(2^{+})~
\phantom{{}^{{}^{\displaystyle{i}}}}\!\!${{{D${}_{{s2}}^{*}$(2573)}}}  
&\phantom{i}{2571.90$\pm$.8\phantom{0}}\phantom{i}  
&\phantom{i}{2574.92}\phantom{i}

\\
\hline\hline
			\end{tabular}
}}\end{small}\\ \spazio{1.6}
Table V
\end{center}
In Table V we give the
         levels of $Bc,\,Bs,\,Ds$ and of the light $u\bar{d}$ mesons in MeV. For the 
		first three families we have $\sigma=1.111,\,1.111,\,1,227$ GeV/fm and 
		$\alpha=0.3591,\,0.3975,\,0.5344$ respectively. For $u\bar{d}$ family
		$m_u=2.94$ and $m_d=6.1$ MeV, $\sigma=1.34$ GeV/fm and $\alpha=0.656$.
		The $\alpha_S$ curve has a steep increase for low masses \cite{PDG}:
		fixing it at a constant value for the whole family induces a large error
		in the pion mass. The pion experimental mass is reproduced
		by taking  $\alpha=0.99$. At a smaller extent the same can be said for $\rho(770)$. See \cite{RadRep,Bada,GodMoa,PS2} for results
		using different methods.

Although the similarities between the atom and the meson spectroscopy are evident, there
are deep differences for e two cases. In atomic systems
all the relevant physical parameters (masses, coupling constant, anomalous magnetic moments)  are fixed at the measured values and the problem is
completely determined up to the precision, actually very high, at which the radiative effects of the electromagnetic field, the finite dimensions of the particles,
the electroweak unification and so on, give non negligible corrections.   
For Quarkonium models, on the contrary, the fundamental physical parameters are not 
available since the beginning. The
masses of the component quarks, $\sigma$ and $\alpha_S$   must be determined by means of a self-consistent fit on the spectral data,  
calculated with the inclusion of the essential contribution of the Breit term.
Limitations to potential models are due to the asymptotic freedom
at short distances and to the creation of light quarks for increasing energy of the
interaction: attempts have been made to describe these effects
by softening the Coulomb-like potential at the origin and screening the scalar term at 
infinity. Each correction, however, introduces new parameters which are fitted so to enhance the agreement with the experimental data.
We could argue that the effectiveness of a model can roughly be appraised by looking at 
the accuracy and the quantity of data it is able to reproduce with the least number of fitted  parameters. In this sense our covariant wave equation is very effective. We have indeed used the least number of fitted parameters: 
the same constant values of $\sigma$ and $\alpha_S$ within each meson 
family and the quark masses. The `flavor independence' due QCD, implies, moreover, that
the string tension  should be expected to be constant, at least for heavy quarks. The
separate fits for the families $b\bar{b},\,c\bar{c}$ and $b\bar{s}$, yield values of $\sigma$ that turn out to be the same within the limits of the  computational error. 	
Finally we have been able to reproduce also the masses of the light $u\bar{d}$ mesons, for which potential models generally fail: in this case, in addition to relativity, 
the contribution of the Breit term is determinant. Moreover the masses for the $u$ and $d$ quarks produced by the fit are 
small and very close to the current algebra masses, in contrast with the much higher 
values of the constituent masses used for potential models.
For greater detail from a technical and phenomenological point of view we refer to  \cite{GS3,BGS_PR}.

\end{document}